\documentclass[fleqn,usenatbib]{mnras}

\usepackage{newtxtext, newtxmath}

\usepackage[T1]{fontenc}

\DeclareRobustCommand{\VAN}[3]{#2}
\let\VANthebibliography\thebibliography
\def\thebibliography{\DeclareRobustCommand{\VAN}[3]{##3}\VANthebibliography}

\usepackage{graphicx}	
\usepackage{amsmath}	

\newcommand{\lmir}{$L_{\rm{MIR}}$}
\newcommand{\loiii}{$L_{\rm{[O iii]}}$}
\newcommand{\lxb}{$L_{\rm{XRB}}$}
\newcommand{\lx}{$L_{\rm{X}}$}

\newcommand{\mbh}{$M_{\rm{BH}}$}
\newcommand{\mstar}{$M_{\star}$}
\newcommand{\msigma}{$M_{\rm{BH}} - \sigma^*$}
\newcommand{\mmstar}{$M_{\rm{BH}} - M_{\star}$}
\newcommand{\arcsecond}{$''$}
\newcommand{\vel}{$\sigma^*$}
\newcommand{\nh}{$N_{\rm{H}}$}
\newcommand{\rlx}{R$_{\rm{LX}}$}
\def\code#1{\texttt{#1}}

\title[Lowest-mass AGNs in the Bo\"otes Field]{Lowest-mass X-ray selected AGNs in the Bo\"otes Field}

\author[Purohit et al.]{Rujuta A. Purohit,$^{1}$\thanks{rujuta.purohit.24@dartmouth.edu}
Ryan C. Hickox,$^{1}$
Grayson C. Petter$^{1}$
\\
$^{1}$Department of Physics and Astronomy, Dartmouth College, 6127 Wilder Laboratory, Hanover, NH 03755, USA}

\date{Accepted 2025 February 19. Received 2025 February 14; in original form 2024 October 28.}

\pubyear{\the\year{}}

\begin{document}
\label{firstpage}
\pagerange{\pageref{firstpage}--\pageref{lastpage}}
\maketitle

\begin{abstract}
    We present a multi-wavelength analysis of three candidate active galactic nuclei (AGNs) in low-mass galaxies in the Bo\"otes field with the aim of improving constraints on the occupation fraction of low-mass black holes (BHs). Galaxies with low stellar masses ($M_{\star} < 10^{9.5} M_{\odot}$) are particularly interesting hosts for AGNs as they may contain BHs that have not grown significantly since the epoch of their formation in the early Universe. Using archival data from the Chandra X-ray Observatory, we find three X-ray luminous low-mass galaxies and assess whether they host AGNs. We find one of these sources to be variable in the X-ray and compute its X-ray light curve and spectrum. We compute the X-ray, mid-infrared, and [O III] luminosities and compare them to established AGN luminosity relationships in the literature. We then fit various star-forming, dust emission, and AGN templates to the spectral energy distributions (SEDs). The star formation rates estimated from the SED fits are unable to explain the observed X-ray luminosities of the candidates, providing more support for the presence of AGNs. By analysing the deviation from linear relationships between X-ray and mid-infrared luminosities, we find these systems to be obscured (with $\log N_{\rm H}[{\rm {cm^{-2}}}] \sim 22.7, > 25.0$, and $24.4$, respectively). We employ the scaling relationship between BH mass and stellar velocity dispersion to estimate the BH masses as $\sim 10^5 - 10^6 M_{\odot}$ and accreting at Eddington ratios $10^{-2} < \lambda_{\rm Edd} <10^{-1}$. 
\end{abstract}


\section{Introduction}
\label{sec:intro}
It is widely accepted that a supermassive black hole (SMBH, $M_{\text{BH}} \approx 10^6 - 10^{10} M_{\odot} $) lies at the centre of most local massive galaxies \citep[e.g.,][]{Magorrian1998, Marconi2003, KormendyHo2013}. The majority of SMBHs lie dormant, but a small fraction produce significant amounts of light and energy, often outshining their host galaxy across the electromagnetic spectrum. These are known as active galactic nuclei \citep[AGN; for reviews see,][]{Salpeter1964, Lynden1969, Shakura1973, Rees1984, Urry1995, Padovani2017, HickoxAlexander2018}.

There are many theories for the formation of SMBHs \citep[e.g.,][]{Shankar2013, Miller2015, Greene20}: the collapse of primordial gas \citep[``heavy" seeds of $\sim 10^4 - 10^5 M_{\odot}$; e.g.,][]{Loeb1994, Bromm2003, Begelman2006, Lodato2006, Johnson2013, Ferrara2014}, super-Eddington accretion of Population III star black hole (BH) remnants in the early-Universe \citep[``light" seeds of $\sim 10^2 M_{\odot}$; e.g.,][]{Madau2001, Whalen2012, Taylor2014, Jiang2019}, and models that involve a gravitational runaway effect \citep[seeds of \mbh $\sim 10^3 - 10^4 M_{\odot}$ e.g.,][]{Portegies2002, Miller2002, Purohit2024}. The challenge to testing these theories is that they require extensive multi-wavelength observations of high-redshift galaxies. Such detailed observations of early-Universe systems were not possible until the recent advent of the James Webb Space Telescope \citep[e.g.,][]{Volonteri2016, Vito2018, Larson2023, Barro2023, Greene2023}. Recent detections of high-redshift objects like ``little red dots" \citep[LRDs; e.g.,][]{Kocevski2023, Harikane2023, Onoue2023, Larson2023, Barro2023, Greene2023, Matthee2023, Kocevski2024} are helping place constraints on the models of BH-galaxy co-evolution.

Studying BHs in the local Universe is a complementary method to probe BH seeds and BH-galaxy co-evolution. Local systems allow us in principle to place constraints on the nature of BH seeds \citep[e.g.,][]{Wassenhove2010, Ricarte2018}, tighten local scaling relationships \citep[e.g.,][]{Pacucci2023}, and differentiate between the possible formation paths \citep[although this has proven to be challenging, e.g.,][]{Volonteri2009, Bellovary2019}.

Currently, only one BH formation channel is well-established: the collapse of a stellar core following a supernova at the end of a massive ($>8 M_{\odot}$) star's evolution. These BHs have masses of the order of tens of $M_{\odot}$ \citep[e.g.,][]{Celotti1999, Remillard2006, Bambi2020}, with an upper limit of $50 \, M_{\odot}$ imposed by pulsational pair instability \citep[][]{Fowler1964, Barkat1967, Rakavy1967, Heger2003, Woosley2017}. BHs with masses $>50-100 M_{\odot}$ have been detected with gravitational wave instruments such as LIGO \citep{Abbott2020}. Many studies have presented the possibility that these BHs are merger remnants of smaller BHs produced through one (or more) of the seeding mechanisms. Nevertheless, if BHs form through stellar collapse, there must be some mechanism by which they grow into the SMBHs we observe today, implying that BHs in an intermediate mass range had to exist. These BHs in the mass gap between stellar mass and SMBHs are called intermediate-mass black holes \citep[IMBHs; for a review, see][]{Greene20}. IMBHs are excellent systems to study as their mass and Eddington ratios may reveal the history of BH growth over cosmic time.

Finding BHs at the nuclei of low-mass galaxies and characterizing their nature is of interest for many reasons, such as finding clues about the origins of SMBHs and probing BH feedback mechanisms in low-mass systems. Because of the steep rise in the galaxy stellar mass function at low masses, even a small occupation fraction of BHs in low-mass galaxies can change the BH density in that regime greatly, making these systems a particularly important component of the BH growth census. Additionally, the quiet merger history of low-mass galaxies, particularly isolated ones, make them good indicators of seeding mechanisms \citep{Volonteri2008}. However, one of the biggest hurdles to studying these systems has been detecting them in multiple wavelength bands \citep[e.g.,][]{Nguyen2018, Baldassare2020}. Many local accreting systems have been revealed using X-ray activity, optical narrow emission line diagnostics to identify AGNs and broad emission lines to estimate BH masses \citep[e.g.,][]{Greene2004, Greene2007, Reines2011, Reines2013, Reines2014, Moran2014, Secrest2015, Pardo2016, Mezcua2016}. Optical variability has also detected many additional such systems \citep{Geha2003, Schmidt2010, Sesar2007, Palanque2011, Choi2014, Baldassare2018, Baldassare2020a}. AGN activity is common in the local Universe, with $\sim 5 - 10\%$ of nearby galaxies found to host optical AGN activity \citep[i.e., identified as Seyfert galaxies from their optical emission-line properties; e.g.,][]{Veilleux1987, Maiolino1995, Ho1997, Kewley2001, Kewley2006, Hao2005}. These observational techniques, among many others, have significantly improved our understanding of the presence of AGNs and accreting BHs in the local Universe. 

Through analyses of SMBHs and their host galaxies, many scaling relations have been developed. One such scaling relation is between the BH mass and the host galaxy's total stellar mass \citep[e.g.,][]{Mancini2012, Reines2015}. Since we know that the SMBH mass correlates with the galaxy mass, we might expect the same for IMBHs found in smaller galaxies. This allows us to essentially verify whether dwarf galaxies ($M_{\star} < 10^9 M_{\odot}$) host low-mass BHs. While there are no dynamically confirmed IMBHs in or not in the nuclei of dwarf galaxies, recent studies have produced a number of promising candidates \citep[e.g.,][]{Neumayer2012, Arcavi2014, Blagorodnova2017, Nguyen2018, Mezcua2018b,Mezcua2018a, Wevers2019, Nguyen2019, Reines2020, 2023Ansh}. 

Low-mass BHs and AGNs found in dwarf galaxies would be crucial for constraining BH seed formation models and better understanding the distribution of BH mass throughout the Universe \citep[e.g.,][]{Pacucci2021, Wasleske2022}. Dwarf galaxies at low redshift may host BHs of similar mass to those in early-Universe galaxies and so provide a probe of their early formation \citep[e.g.,][]{Greene20, Latimer2021}. In the local Universe, it is estimated that $\sim80\%$ of nearby, low-mass early-type galaxies host nuclear BHs \citep{Nguyen2018}, but the total population of BHs in low-mass galaxies remains poorly constrained. This is significant, considering the first two prototype low-mass AGNs hosting IMBHs were detected in the local Universe \citep[NGC 4395;][]{Filippenko1989} and \citep[POX 52;][]{Kunth1987, Barth2004}. Thus, searching for AGNs in dwarf galaxies is a valuable method of identifying low-mass BHs \citep[e.g.,][]{Reines2013, Greene20}. Apart from being located at the centers of dwarf galaxies, some IMBHs can be found ``wandering" around their host galaxies such as HLX-1 \citep{Farrell2009} and in massive star clusters such as 3XMM J215022.4-055108 \citep{Lin2020}.

X-ray observations are effective at identifying AGN activity due to their high contrast with the host galaxy and strong penetrating power \citep[e.g.][]{Brandt2015}. However, even X-ray-selected samples suffer some bias against obscured sources or those with substantial X-ray emission from stellar processes. Thus, we use X-ray observations to guide our primary search for targets and then follow up with multi-wavelength data by studying relationships between luminosities at different wavelengths and host galaxy properties (such as the stellar mass, \mstar, star formation rate (SFR), and stellar bulge velocity dispersion, \vel.

This paper is organized as follows: \S \ref{sec:data} details our search and the sample of targets studied. In \S \ref{sec:analysis}, we compare the observed properties of our targets to established relationships in literature. We carry out a spectral energy distribution (SED) template fitting and analyse the mid-infrared (MIR) and [O III] luminosities. We also study the X-ray variability of our sources. In \S \ref{sec:mass_estimates}, we estimate the masses of the central BHs through different scaling relations and calculate the Eddington ratios to better interpret our findings. \S \ref{sec:conclusion} summarizes our results and details the scope for future work.

In this work, we adopt a ``Planck 2015" $\Lambda-$CDM cosmology \citep{Planck2016}: $H_0 = 67.8$ km s$^{-1}$ Mpc$^{-1}$, $\Omega_{\text{m}} = 0.308$, and $\Omega_{\Lambda} = 0.692$. Logarithmic values are base 10 unless otherwise noted. All uncertainties presented are 1$\sigma$ or 68\% confidence intervals unless mentioned otherwise.

\section{Data}
\label{sec:data}
\begin{figure*}
    \centering
    \includegraphics[width=0.41\textwidth]{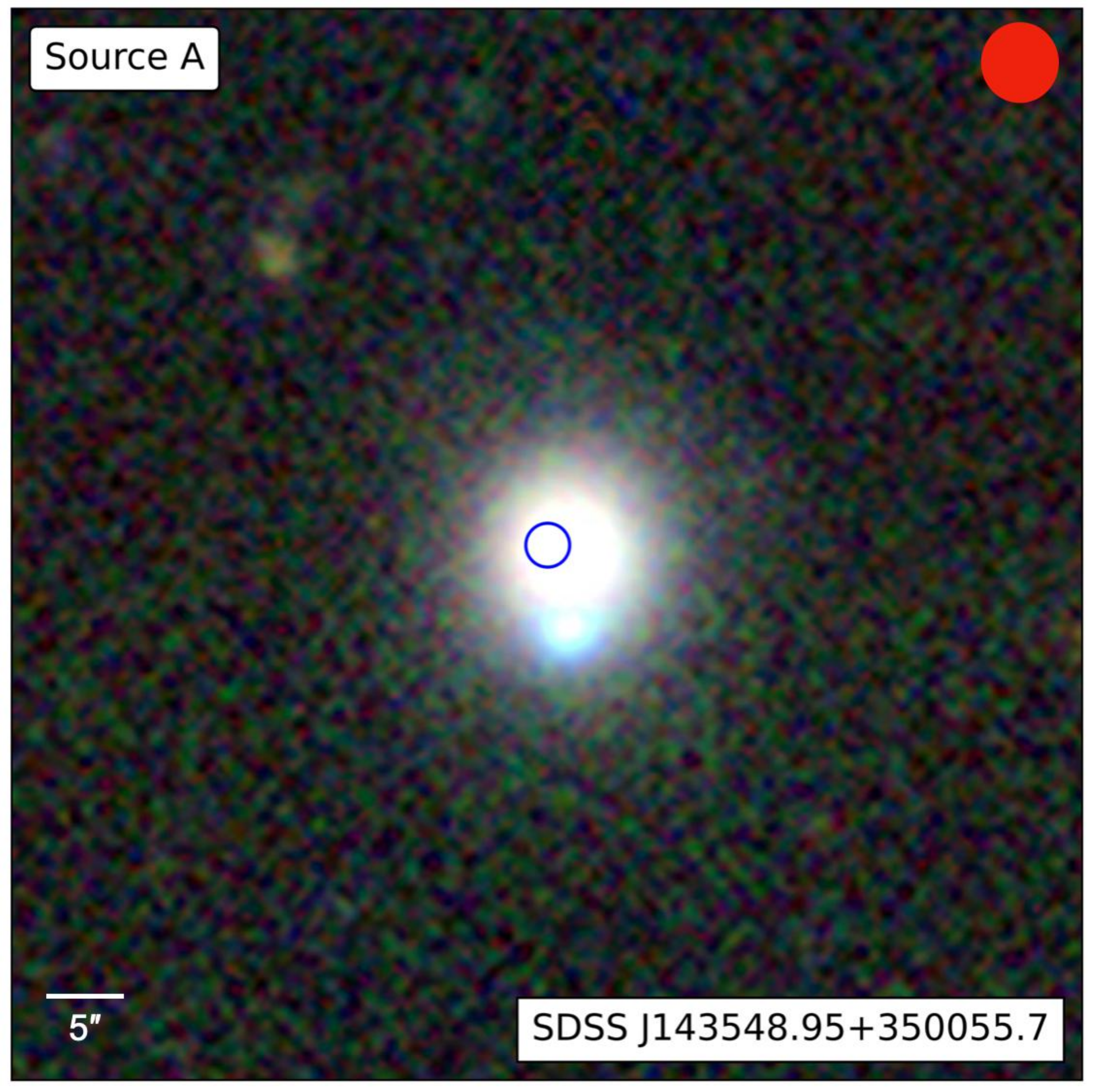}
    \includegraphics[width=0.41\textwidth]{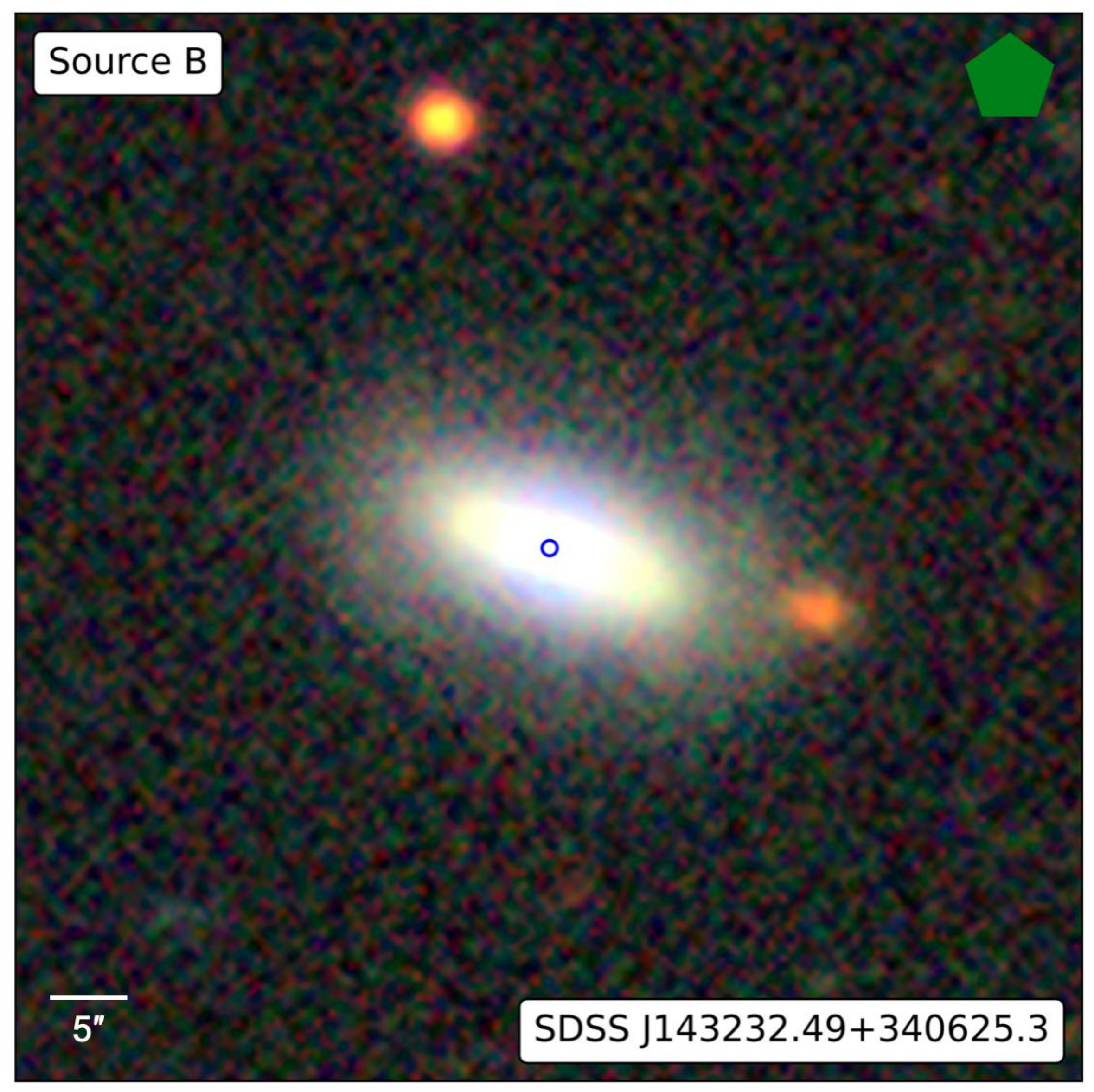}
    \includegraphics[width=0.41\textwidth]{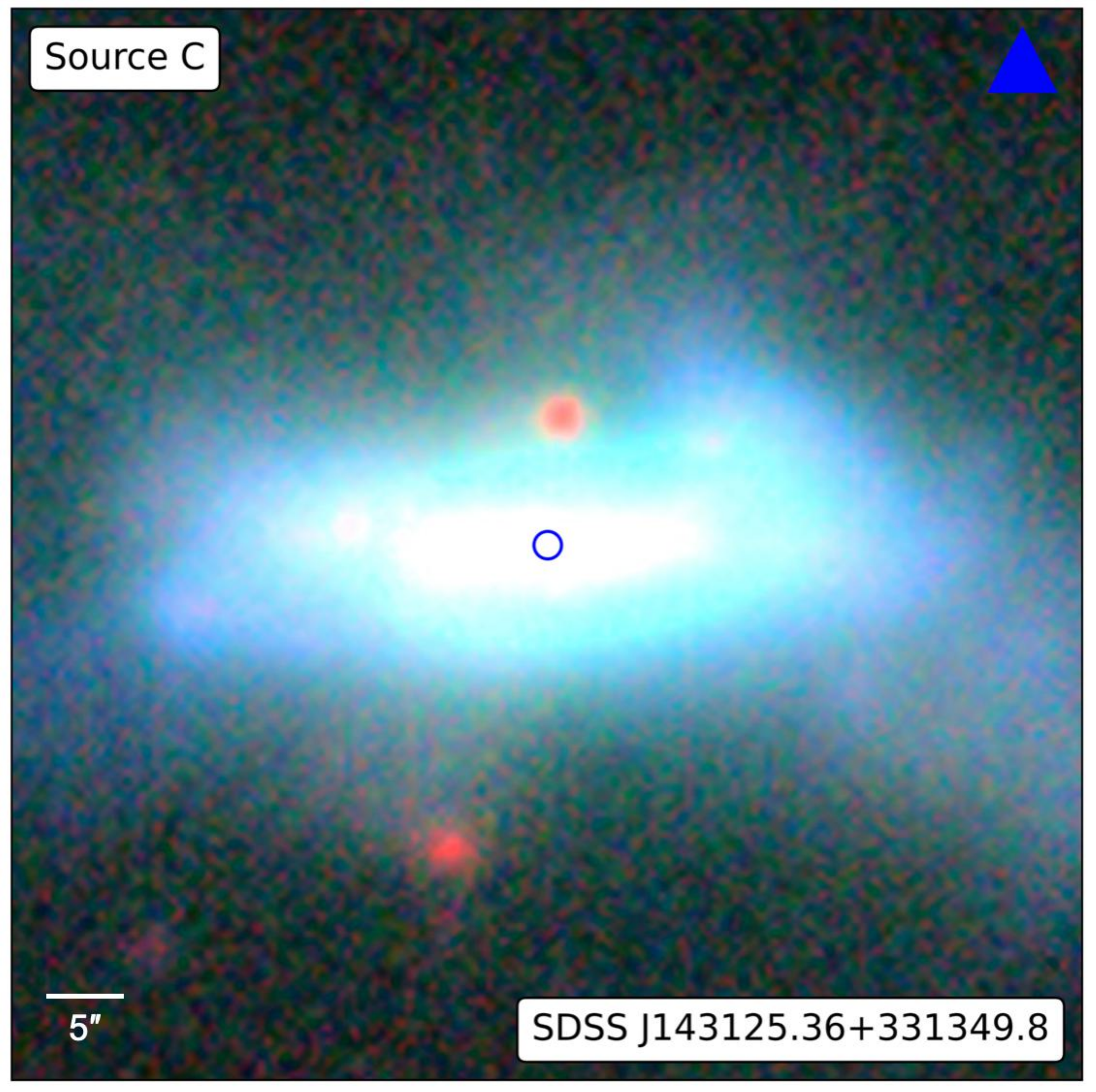}
    \caption{Optical-NIR (\textit{g, r, z}) colour images of our sample of local ($z < 0.05$) low-mass galaxies ($M_{\star} < 10^{9.5} M_{\odot}$) in the Boötes field with candidate X-ray AGN. The images are obtained from the DESI Legacy Sky Survey Browser. The object name of each source is listed in the bottom right corner, the identifier in the top left corner, and the chosen icon in the top right corner.  Blue circles are centred on the coordinates of the X-ray source and the radii of the circles are equal to the 95\% positional errors calculated using Eq 12 in \citet{Kim2007}. The scale is shown in the lower left corner. \textit{Image credit}: Legacy Surveys / D. Lang (Perimeter Institute)}
    \label{fig: images}
\end{figure*}

\begin{table*}
\centering
\caption{Physical properties of the targets studied in this work. X-ray luminosity is reported for the 0.5-7 keV band.}
\label{tab: data}
    \begin{tabular}{cccccccc}
    \hline 
    Object Name & Identifier & Redshift & Chandra exposure & Positional uncertainty & $\log$ \lx & $\log$ \lmir & $\log$ \loiii\\
     &  &  & {time (s)} & (arcsecond) & (erg s$^{-1}$) & {(erg s$^{-1}$)} & (erg s$^{-1}$) \\
     \hline
    SDSS J143548.95+350055.7 & A & $0.0285$ & $28199.4$ & 0.86 & $40.1^{+ 0.2}_{- 0.1}$ & $40.1$ &$ 39.1$ \\
    SDSS J143232.49+340625.3 & B & $0.0423$ & $45452.4$ & 0.31 & $41.1^{+0.1}_{-0.1}$ & $43.6$ & $40.7$\\
    SDSS J143125.36+331349.8 & C & $0.0226$ & $22072.8$ & 0.54 & $40.3^{+0.2}_{-0.1}$ & $42.1$ & $40.7$\\
    \hline
    \end{tabular}
\end{table*}

\begin{figure}
    \includegraphics[width=\columnwidth]{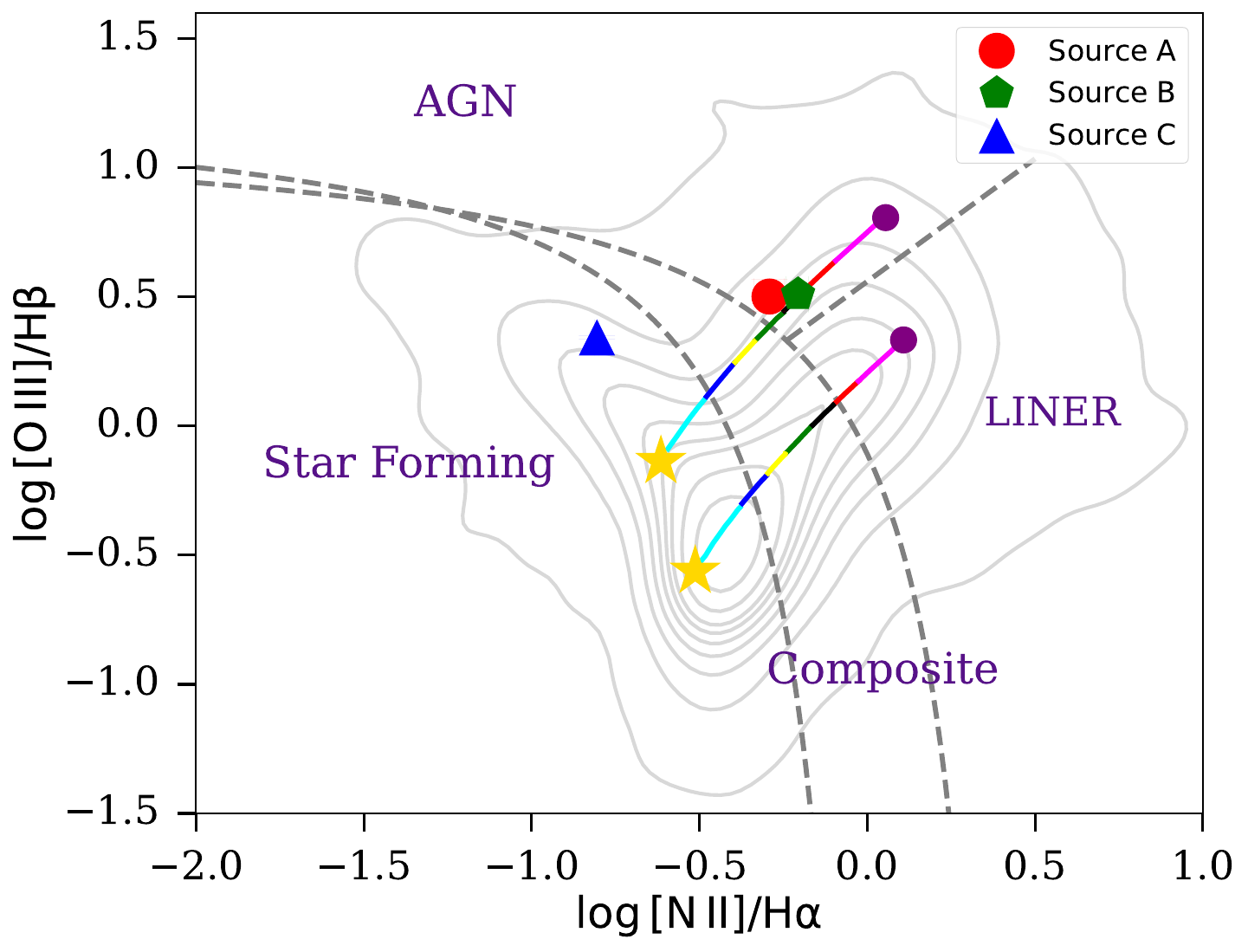}
    \caption{The three low-mass galaxies we study in this work are placed on a narrow-line BPT diagnostic diagram \citep{BPT1981} with demarcations as determined by \citet{Kewley2006}. Two of the targets (A and B) lie in the AGN-demarcated region while source C lies slightly above the nominal star-forming galaxy locus, and this is consistent with a non-negligible contribution ($\sim 30\%$) of the AGN to the [O iii] luminosity calculated using \citet{Jones2016}. We also plot the tracks from \citet{Kauffmann2009} used to estimate AGN contribution based on distance from the ``pure" star-forming sequence (gold star) to the ``pure” AGN region (purple circle). The colour scheme follows from \citet{Kauffmann2009} and \citet{Jones2016}.}
    \label{fig: BPT}
\end{figure}

Our approach to identifying AGN candidates is to search for X-ray counterparts to low-mass galaxies, as luminous X-ray emission is an unambiguous indicator of AGN activity in a galaxy \citep[e.g.,][]{Miller2015, Pardo2016, Baldassare2015}. This analysis requires deep and overlapping optical and X-ray data as low-mass galaxies are expected to have intrinsically low luminosities. The X-ray waveband identifies AGNs while the optical provides initial estimates of redshift and stellar mass. The analysis would also benefit from panchromatic photometry to study the host galaxy and AGN properties through spectral energy distribution modelling. The 9.3 deg$^2$ Bo\"otes extragalactic survey field satisfies all of these requirements because of the investment of X-ray to far-infrared photometry, spectroscopy, and analysis. We utilize two catalogues in this work: the Chandra Deep Wide-Field Survey (CDWFS) in the Bo\"{o}tes Field \citet{Masini2020} (hereafter M20) for X-ray source identification, and \citet{Duncan2021} (hereafter D21) for photometric redshift and stellar mass values. Additional data such as optical and IR photometry is obtained from NASA/IPAC Extragalactic Database (NED) \footnote{https://ned.ipac.caltech.edu/} and allows us to calculate initial estimates of the sources' luminosities, distances, and star formation rates. We also obtained all Chandra observations covering our selected sources (including some later observations that came after the production of the CDWFS catalogue) from the Chandra Data Archive at Webchaser\footnote{https://cda.harvard.edu/chaser/}.

We matched the CDWFS catalogue of the X-ray detections in M20 with the optical-IR positions of galaxies in D21 with a matching radius of 1". D21 contains stellar mass estimates (estimated using a Python-based code described in \citealt{Duncan2019}) which we rely on to identify low-mass galaxies. M20 contains 6891 sources while D21 has 602,949 sources in the Bo\"otes field. Since we require optical-IR data for our analysis, we maintained only those sources in our sample that had reliable multi-wavelength photometry. We limited the sample to $z < 0.05$, making redshift cuts using the values reported in D21 which are a combination of photometric and spectroscopic values. For the Bo\"otes field, the spectroscopic redshifts are primarily obtained from the AGN and Galaxy Evolution Survey \citep[AGES;][]{Kochanek12}. The photometric redshift estimates are obtained through SED fits and are superseded by the spectroscopic ones whenever applicable. We selected only those sources that had reliable optical counterparts. After these initial cuts, our sample had 168 total sources.

We selected low-mass galaxies by choosing only those candidates with $M_{\star} < 10^{9.5} M_{\odot}$. The stellar masses in D21 are derived using a grid-based SED fitting approach that scales to the large samples available across the Bo\"otes field. For nearby galaxies, \mstar \ might be underestimated in D21 because the method uses 3" \, apertures for optical to NIR bands and 4" apertures for the IR. This aperture measures a smaller fraction of light from large sources like nearby galaxies (see D21 Section 2.1 for more details). Thus, due to the choice of the aperture sizes, \mstar \, estimated through aperture photometry will be systematically underestimated for local objects. We point out that we re-compute stellar masses for our sources using multi-wavelength SED fitting (see \S \ref{sec: sed}). Previous cuts made for optical spectra allowed us to examine our low-mass galaxy sample using velocity dispersion estimates. Any objects with \vel $> 150$ km s$^{-1}$ are likely to have a stellar mass larger than reported ($M_{\star} > 10^{10} M_{\odot}$) and are not likely to host lower-mass BHs. This is based on the \msigma \, relation in \citet{Greene20}, where we estimate \vel \, required for \mbh $> 10^7 M_{\odot}$, considering a generous upper limit on the term ''low-mass BH". We found 14 sources with \vel \, $< 150$ km s$^{-1}$. After removing these objects, we were left with three sources that we analysed in this paper.

We calculated the X-ray luminosities (F band, 0.5–7.0 keV) of the sources using flux values listed in the Chandra Deep Wide-Field Survey in the Boötes field catalogue in M20. The luminosities of the three targets, along with their coordinates, Chandra exposure times, redshifts, and D21 reported masses are listed in Table \ref{tab: data}.

We obtained optical-NIR images from the DESI Legacy Sky Survey Browser\footnote{https://www.legacysurvey.org}. The images are shown in Figure \ref{fig: images}. The identification is displayed as a circle centred at the coordinates of the X-ray source with a radius equal to the 95\% positional uncertainty as calculated using Eq. 12 in \citet{Kim2007}:

\begin{equation}
\log \text{PU} = \\ 
\begin{cases}
    \begin{aligned}
    & 0.1145 \, \text{OAA} - 0.4958 \log{\rm C} + 0.1932, \\
    & \quad 0.0000 < \log C \leq 2.1393,
    \end{aligned} \\
    \begin{aligned}
    & 0.0968 \, \text{OAA} - 0.2064 \log{\rm C} - 0.4260, \\
    & \quad 2.1393 < \log{\rm C} \leq 3.3000.
    \end{aligned}
\end{cases}
\end{equation}

where PU is the positional uncertainty in arcseconds, off-axis angle OAA is in arcminutes, and C is source counts. To account for the newest observation for Source B (Obs 18475) which is not included in M20, we use CIAO tools such as \code{celldetect} (because of its accuracy in detecting faint point sources) to obtain the number of counts. After correcting for the astrometric offset in M20 (in RA by -0.27462", DEC offset by -0.06042"), we obtain the PUs with a 95\% confidence.

We analysed the optical emission from the nuclear region of the sources using single-exposure images from the Legacy Surveys DR9. For each galaxy, the brightest optical emission is located at the X-ray position, well within the 95\% PU of the X-ray source.

\begin{figure}
\centering 
    \includegraphics[width=\columnwidth, keepaspectratio]{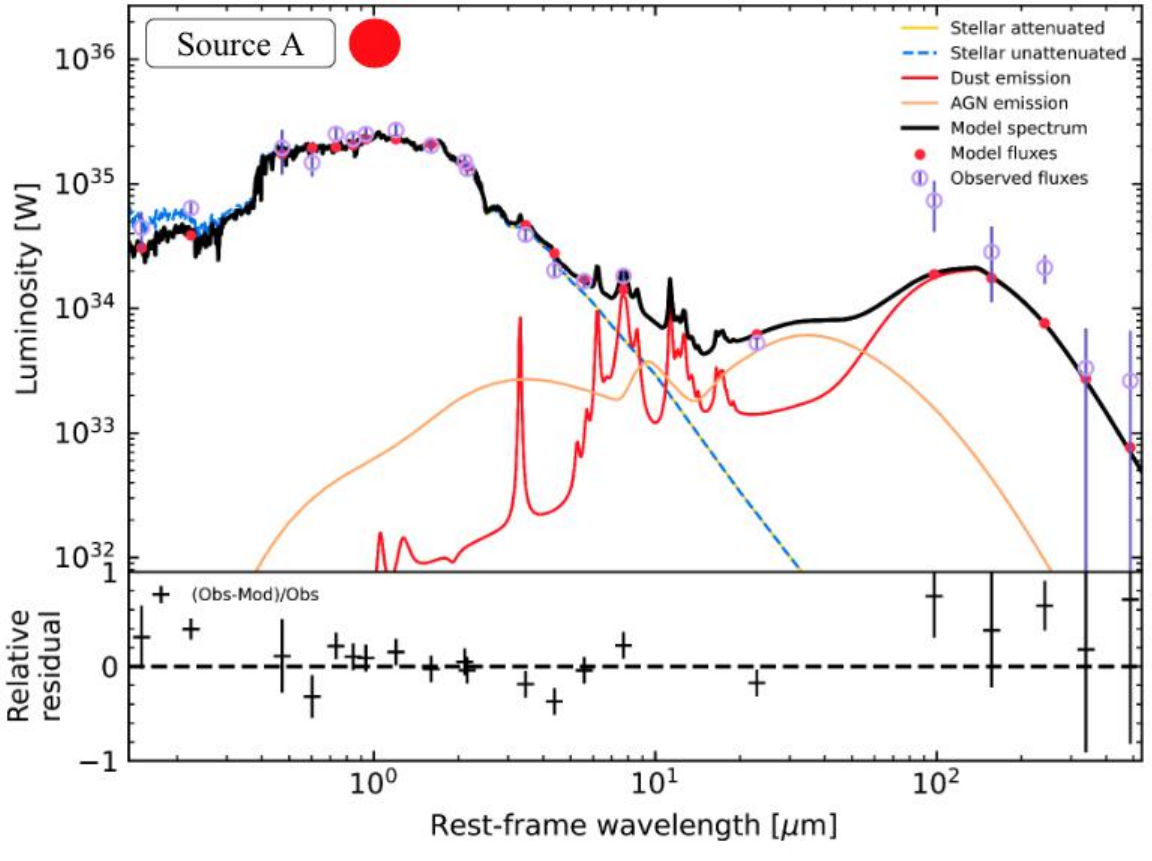}
    \includegraphics[width=\columnwidth, keepaspectratio]{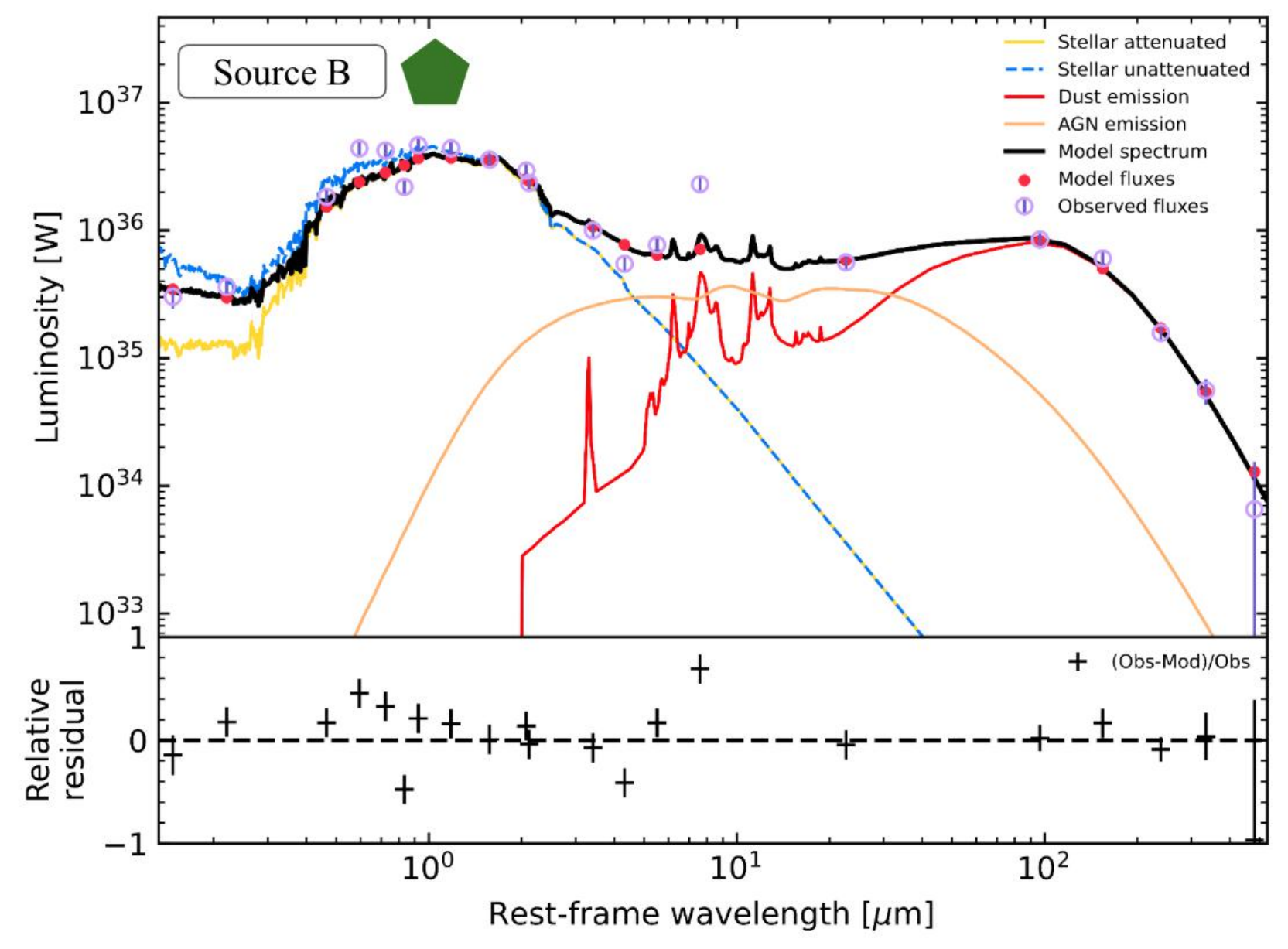}
    \includegraphics[width=\columnwidth, keepaspectratio]{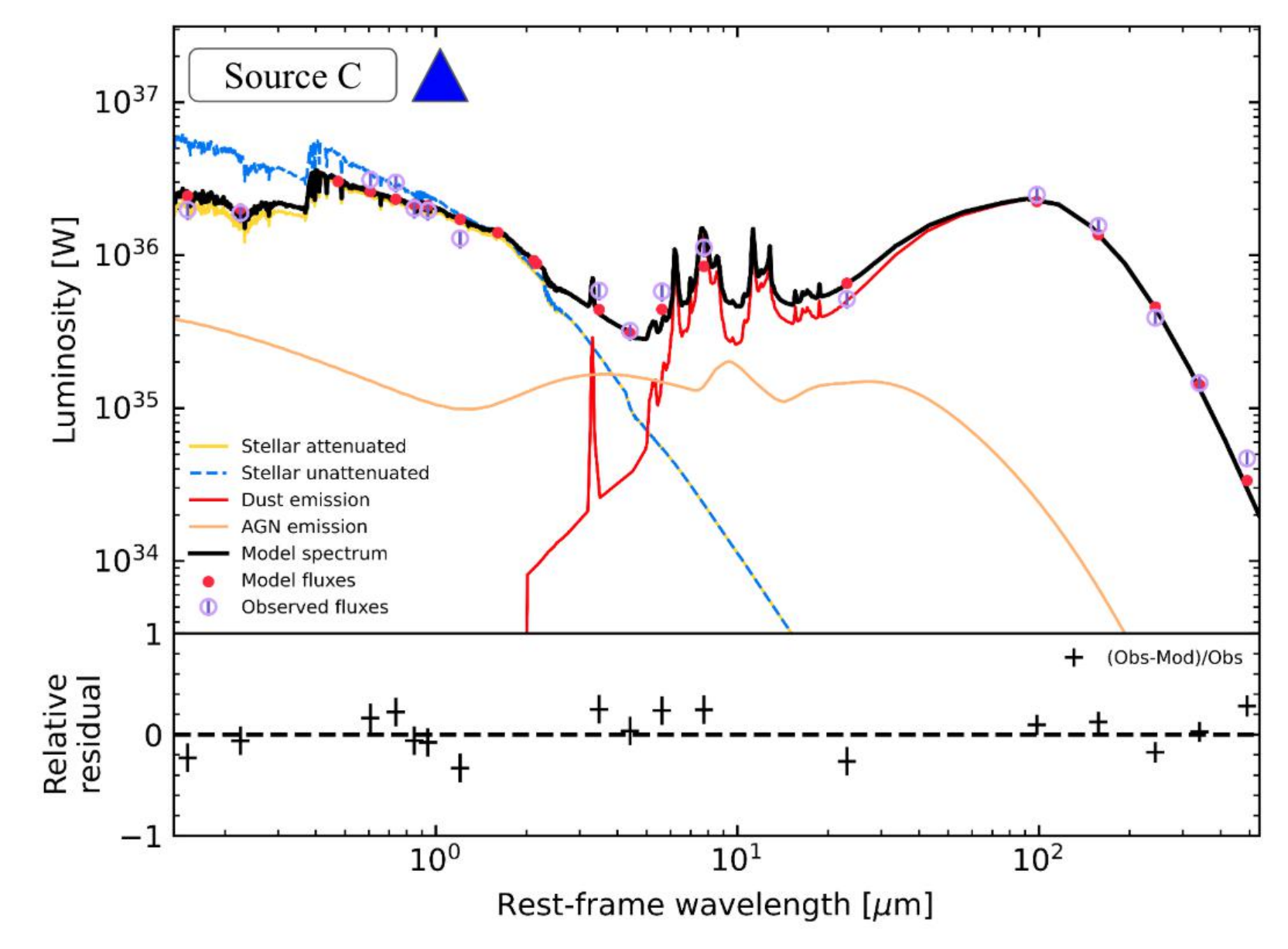}
    \caption{The results of SED fitting using photometric data done in CIGALE. We plot luminosity (W) on the y-axis and the rest-frame wavelength on the x-axis (microns). The different curves represent the templates used in the fitting such as the star formation history, AGN structure, and dust attenuation. The relative residuals are plotted in the bottom panels of the three figures.}
    \label{fig: SEDs}
\end{figure}

We consider the chance probability of the coincidence of the X-ray sources with the optical centres of the host galaxies. With our matching radius of 1\arcsecond \, and the estimated number of low-mass galaxies in the Bo\"otes field, we find that the probability that our sources are background objects is $\lesssim 2 \%$. Furthermore, we also check this with the entire 9.3 deg$^2$ field and the 6981 X-ray sources in M20. We find similar results with a probability of $\lesssim 1.8 \%$. 

The optical narrow-line emission observed in the Sloan Digital Sky Survey (SDSS) spectra also provides a useful check for identifying these sources as AGNs. Figure ~\ref{fig: BPT} shows a narrow-line [N II]/H$\alpha$ vs. [O III]/H$\beta$ Baldwin, Phillips \& Terlevich (BPT) diagnostic diagram \citep{BPT1981} with demarcations determined by \citet{Kewley2006}. Sources A and B lie in the AGN demarcated region while C lies in the star-forming region. We include source C in our sample due to its SED fitting revealing an AGN contribution and the X-ray source overlapping with the optical centre. We estimate the AGN contribution to source C using its position on the BPT diagram (as discussed in detail in \S \ref{sec: lum}), finding it to be  $\sim 30\%$, although with considerable uncertainties given the width of the star-forming sequence on the BPT diagram. In Figure \ref{fig: BPT}, we also plot the tracks from \citet{Kauffmann2009} used to determine the AGN contribution relative to the position of the source from the pure star-forming (gold star) and pure AGN points (purple circle).

\section{Analysis and Results}
\label{sec:analysis}
\subsection{Spectral Energy Distribution (SED)}
\label{sec: sed}

To investigate the nature of our sources, we begin by creating spectral energy distributions (SEDs) for each source. SED fitting decomposes integrated galactic emission into its components and allows us to infer physical properties such as \mstar, SFR, and AGN luminosity \citep[e.g.,][]{Magris2015, Yamada2023}.

We utilize SED fitting with photometric fluxes at a range of wavelengths obtained from NED. The photometry used come from GALEX \citep{GALEX} for the UV (NUV and FUV bands), SDSS DR17 \citep{Abdurro'uf2022} for the optical-NIR, and the Spitzer Space Telescope \citep{Spitzer-IRAC} for the MIR. In the far-IR, we use data from the Herschel Space Telescope's Photodetector Array Camera and Spectrometer (PACS) and Spectral and Photometric Imaging Receiver (SPIRE), as compiled in the Herschel Extragalactic Legacy Project catalogue \citep[HELP;][]{HELP}. In the IR, we choose to use Spitzer instead of the Wide-field Infrared Survey Explorer because of the higher sensitivity of the former in the mid-IR wavebands. Specifically, we use photometric measurements values from integrated maps, ensuring that all the light is included. For GALEX and 2MASS, we use the Kron aperture fluxes; for SDSS, we use the CModel values. For the Spitzer/IRAC data, all reported photometry is from a fixed aperture. We use flux values from the 5.8" aperture from Spitzer but we note that this is within the errors from WISE which uses larger apertures and flux integrated maps. Visual inspection also confirms a smooth SED shape, indicating consistency across the different datasets and reinforcing the reliability of our photometric measurements for SED fitting and analysis.

\begin{table*}
    \centering
    \caption{\centering Obscuration of sources based on luminosity in 0.5-7 kev Chandra band and the SED-derived MIR luminosity.}
    \label{table: obscuration}
\begin{tabular}{cccccc}
    \hline 
    Identifier & log \lx & log $L_{\rm{X}}(L_{\rm{MIR}})$  & R$_{\rm{LX}}$ & log \nh & HR  \\
    & (erg s$^{-1}$) &  (erg s$^{-1}$)  & & (cm$^{-2}$) & \\
    \hline
    A & $40.1^{+ 0.2}_{- 0.1}$ & $40.5$ &  $-0.38$ & $22.71^{+0.5}_{-0.3}$ & $-0.54^{+0.2}_{-0.5}$\\
    B & $41.1^{+0.1}_{-0.1}$ & $43.4$ &  $-2.32$ & $> 25.0$ & $-0.59^{+0.08}_{-0.09}$\\
    C & $40.3^{+0.2}_{-0.1}$ & $42.2 $ &  $-1.86$ & $24.4^{+0.2}_{-0.2}$ &   $-0.14^{+0.3}_{-0.2}$ \\
    \hline 
\end{tabular}
\end{table*}

We use the Code Investigating GALaxy Emission \citep[CIAGLE;][]{Burgarella2005, Boquien2019} to fit the SED to different galaxy types which gives us important initial insights into the spectral properties of these galaxies and allows us to estimate the AGN luminosity. In our fitting, we use the Salpeter initial mass function \citep{Salpeter1955}. Using the Chabrier IMF \citep{Chabrier2003} introduces a systematic offset of 0.3 dex between the stellar mass estimates. Within CIGALE, we utilize the AGN torus model \code{skirtor2016} \citep{Stalevski2016}, dust emission model \code{dale2014} \citep{Dale2014}, the attenuation model \code{dustatt\_calzleit} \citep{Calzetti2000}, and various star formation history models. We compute the best-fit models using the \code{pdf\_analysis} method. The best-fit parameters from our models are listed in Table \ref{tab: X-ray binary}. The CIGALE fit to the SED returns stellar masses, star formation rates, and other physical quantities, along with their uncertainties. For Source A, we should note that the best-fit model is slightly low relative to the data in the FIR component, which means the SFR could be higher. 

Using the SEDs, we estimate the AGN contribution to the MIR luminosity, \lmir, by interpolating the luminosity of the AGN component to its monochromatic value of $\nu L_\nu$ $6\mu$m since this is used by \citet{Chen2017} in their analysis.

To check our results from CIGALE, we also carry out a simple \code{scipy} SED fitting using the templates from the low-resolution templates from \citet{Assef10}. The templates in \citet{Assef10} are generated using empirical data from the multi-wavelength photometric observations of the NDWFS \citep{Jannuzi1999} and the spectroscopic observations of AGES \citep{Kochanek12} which cover the Boötes field extensively. We find that these two techniques are in good qualitative agreement.

\subsection{Luminosity Relations}
\label{sec: lum}

\begin{figure}
    \centering
    \includegraphics[width=\columnwidth, keepaspectratio]{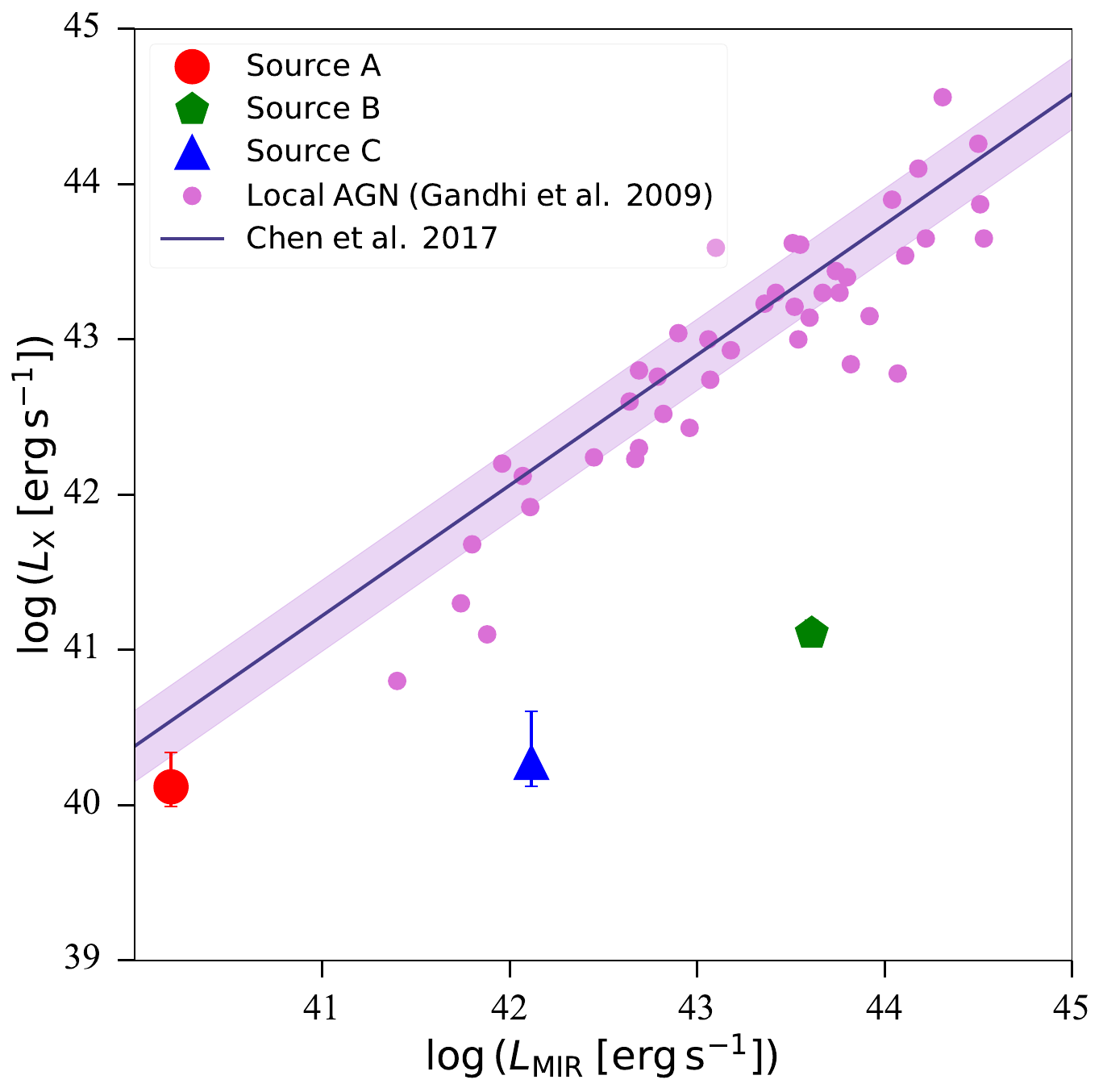}
    \caption{A plot of the derived MIR vs the X-ray luminosities of the 3 galaxies along with the integrated scaling relation in \citet{Chen2017}. The MIR luminosities are derived by integrating over the AGN components of the SED fitting. We also plot a sample of local AGN from \citet{Gandhi2009} (solid pink circles) for reference. The three sources fall in the low-luminosity tail of the scatter. The deviation from the linear relationships suggests significant obscuration.}
    \label{fig: LXvsLMIR}
\end{figure}

In this section, we explore the empirical relationship between the X-ray and MIR luminosities of AGN \citep[e.g.,][]{Gandhi2009, Stern2015}. We obtain the MIR luminosity from the SED fitting, as described in Section \ref{sec: sed}. We use the integrated scaling relation from \citet{Chen2017} to compare the X-ray and MIR emissions. Along with our targets, we also plot local AGN as listed in \citet{Gandhi2009} to compare our sample of targets to previous studies. This is shown in Figure \ref{fig: LXvsLMIR}.

Following \citet{Chen2017}, we calculate the $6\mu$m MIR luminosity for our sources and compare them against the X-ray. The intrinsic dispersion in the relationship is 0.23 dex \citep{Carroll2023}. We see that our targets are found at the lower end of the trend: their MIR luminosities are consistent with those of other AGNs but the significant deviation ($ \geq 1$ dex) from the lowest X-ray luminosities suggests significant obscuration. The deviation from the linear relationship can be used to study the AGNs' obscuration \citep[e.g.,][]{Carroll2021, Carroll2023}. While these sources each have too few X-ray counts (15, 104, and 28 counts, respectively) to perform detailed spectral fitting, we can get a rough sense of their spectral shapes via their hardness ratios. The hardness ratios in M20 (HR = (H - S)/(H + S), where H are hard X-ray counts in the energy range 2.0–7.0 keV and S are soft X-ray counts in the energy range 0.5–2.0 keV) are computed using the Bayesian Estimator for Hardness Ratio \citep[BEHR, ][]{Park2006}. The hardness ratios for sources A, B, and C are HR $=-0.54^{+0.15}_{-0.46}$, $-0.59^{+0.078}_{-0.09}$,  $-0.14^{+0.32}_{-0.25}$ respectively. We estimate the column densities (\nh) following \citet{Carroll2021} and \citet{Carroll2023} that relate \nh \, to the ratio of the observed to expected \lx, \rlx (from the linear relationship in \citet{Chen2017}).  \nh \, values are estimated using parameters of choice from the BORUS model in \citet{Carroll2023}: $\Gamma = 1.9$, $\sigma_{\rm scatt} = -1.04$, R = 0.99, and OA $= 60^{\circ}$. Uncertainties are calculated by bootstrapping over the full range of parameters. Source B has a \rlx \, value that was outside our grid of models, so the \nh \, value was extrapolated and should be thus treated as a lower limit. The column densities and the associated errors are listed in Table \ref{table: obscuration}. 

We note that sources B and C have relatively low values of X-ray hardness ratio $(< -0.14)$ despite having high estimates of \nh \, from their X-ray to MIR luminosity ratios. At low redshift, the observed frame X-ray hardness is expected to become soft at the highest column densities, as the weak scattered component dominates over the heavily obscured direct component \citep[e.g.][]{Brightman2014}; the observed X-ray hardnesses are thus consistent with what may be expected for these heavily obscured AGN.

The scaling relationships between \lx \, and \loiii \, have been widely studied \citep[e.g,][]{Georgantopoulos2010, Bongiorno2010, Yan2011}. The relationships between \loiii \, and \lmir \, are also well defined \citep[e.g,][]{LaMassa2010, Georgantopoulos2010}. The [O iii] luminosity is considered to be a good proxy of nuclear power because it originates from the narrow-line region which is outside the obscuring dusty torus and is often utilized as an isotropic luminosity indicator \citep[e.g.,][]{Mulchaey1994, Bassani1999, Young2001, Schmitt2003, Heckman2005, Georgantopoulos2010, Yan2011, Zhang2017, HickoxAlexander2018}. Together, the [O iii] and intrinsic X-ray luminosities are good indicators for the AGN bolometric luminosity \citep{Heckman2004}. Among Type 1 AGNs, for which there is an unobstructed view of the accretion disk, \lx \, is found to correlate very well with \loiii \, \citep[e.g.,][]{Mulchaey1994, Heckman2005}. We obtain the [O iii] line fluxes from SDSS DR 17 and correct them for extinction.

\begin{figure}
    \centering
    \includegraphics[width=\columnwidth, keepaspectratio]{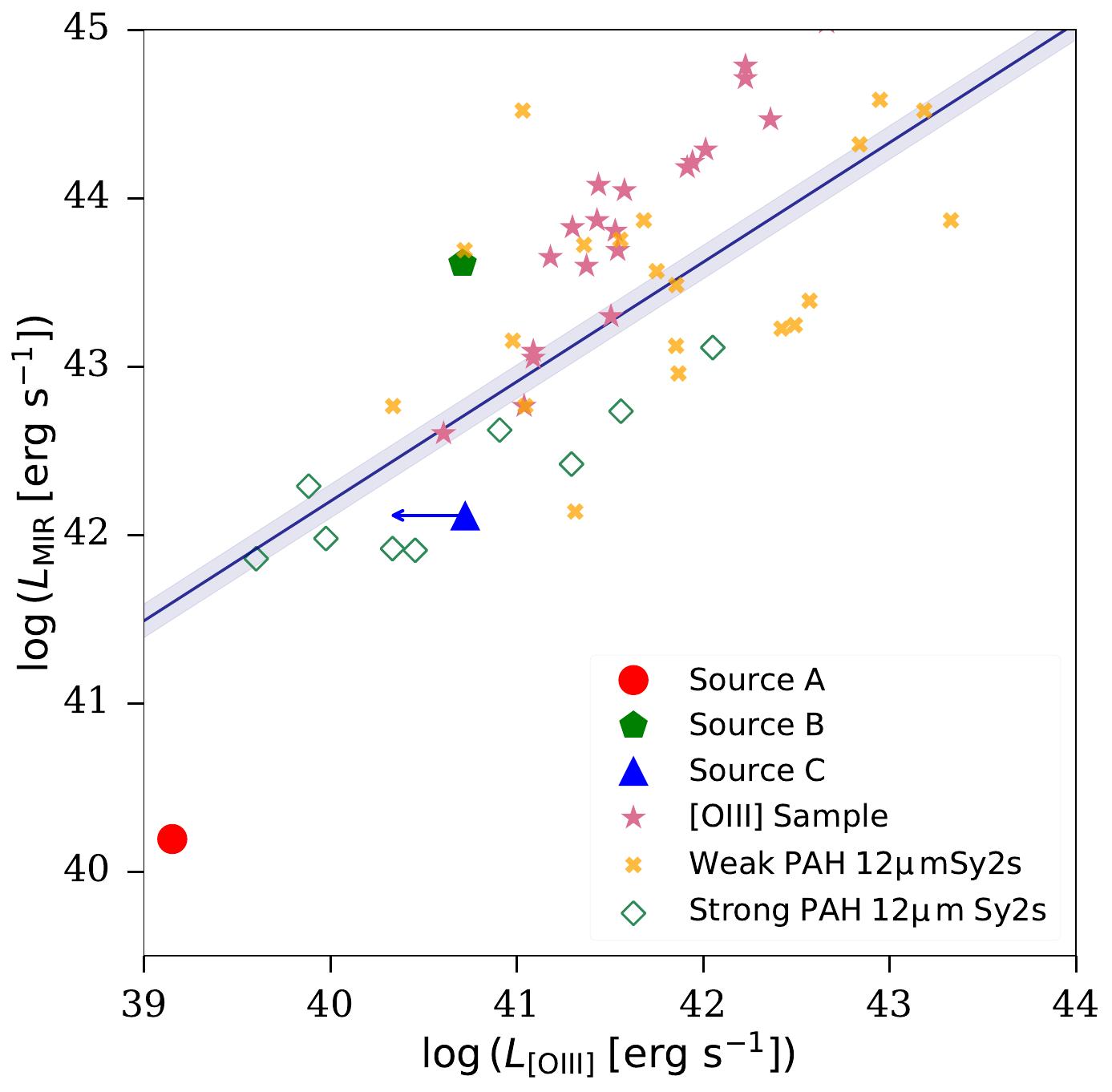}
    \caption{A plot of \loiii \ vs \lmir, along with the relationship derived by \citet{LaMassa2010}. We also plot the samples of [O iii] AGN from \citet{Adelman2006} (solid pink stars) and $12 \mu \text{m}$ samples of weak (solid yellow crosses) and strong (open green diamonds) polycyclic aromatic hydrocarbons (PAHs) Seyfert 2 galaxies from \citet{Spinoglio1989}. Sources B and C fall within the scatter while A lies near the tail of the linear relationship extrapolated to the low-luminosity region. Similar to Figure \ref{fig: LOIII vs LX}, the arrow at source C points from the composite [O iii] luminosity to the AGN-only.}
    \label{fig: loiiivslmir}
\end{figure}

\begin{figure*}
    \centering
    \includegraphics[width=0.70\textwidth]{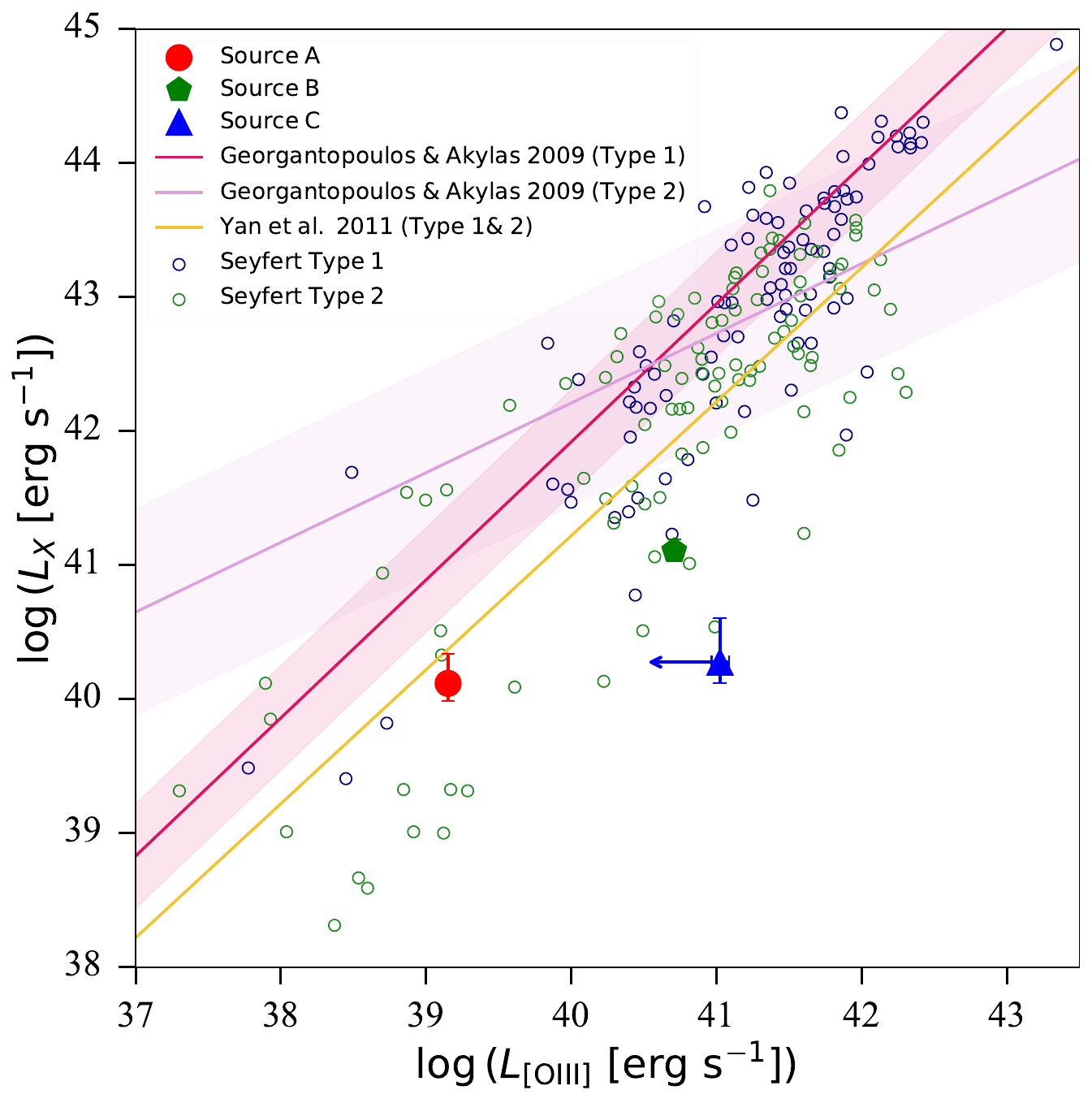}
    \caption{The observed X-ray and [O iii] luminosities of the three objects, plotted with various scaling relations from \citet{Georgantopoulos2010} and \citet{Yan2011} for Seyfert Type 1 and 2 galaxies. We also plot local Seyfert Type 1 (open blue circles) and Type 2 (open green circles) galaxies. The [O iii] luminosity values plotted here are corrected for extinction. We observe that the three sources lie towards the lower luminosity spectrum. The [O iii] is considered to be a good proxy for AGN activity since it originates from the narrow-line region, which is outside the obscuring dusty torus \citep[e.g.,][]{Mulchaey1994, Bassani1999, Young2001, Schmitt2003, Heckman2005, Yan2011, Zhang2017, HickoxAlexander2018}. The [O iii] luminosities are consistent with the expected extrapolation of the samples here. Like the \lx-\lmir \, relation in Figure \ref{fig: LXvsLMIR}, the deviation of the X-ray luminosities below the intrinsic relationship suggests obscuration.  The arrow at source C points from the composite [O iii] luminosity to the AGN-only. The length of the arrow represents the AGN contribution to [O iii] which is $\sim 30\%$.}
    \label{fig: LOIII vs LX}
\end{figure*}

We determine the contribution of the AGN to the galaxy's integrated [O iii] luminosity using the BPT tracks computed by \citet{Jones2016}. The AGN contribution is based on the relative distance in the BPT diagram of the source from pure star-forming and pure AGN regions. Using \citet{Kauffmann2009}, \citet{Jones2016} and Casey et al (in prep)\footnote{https://www.github.com/quinn-casey/f\_agn}, we make estimates of the AGN contribution to [O iii] for each source. Sources A, B, and C have an AGN contribution to the [O iii] luminosity of approximately $90\%, 80\%,$ and $30\%$ respectively. We plot the corrected, AGN-only luminosity for A and B in Figure \ref{fig: LOIII vs LX}. We also plot the composite [O iii] luminosity for source C with an arrow pointing to the AGN-only \loiii.

Next, we examine the relationship between the [O iii] and the MIR as described in \citet{LaMassa2010}. We plot the values of the targets, along with the data sets used in \citet{LaMassa2010} in Figure \ref{fig: loiiivslmir}. We see that the targets fall within the scatter of the relationship as the trend continues to low luminosities.

\begin{table*}
\centering
\caption{Properties derived from CIGALE SED fitting and used to calculate expected emission from X-ray binaries using \citet{Lehmer2019}}
\label{tab: X-ray binary}
    \begin{tabular}{ccccccccc}
    \hline
    Identifier & log \mstar & SFR  & log sSFR & log \lxb & log \lx &  Nuclear light & Nuclear SFR & N($>L_X)$\\
     & ($M_{\odot}$) &  ($M_{\odot}$ yr$^{-1}$) &  (yr$^{-1}$)&  (erg s$^{-1}$) & (erg s$^{-1}$) & fraction & ($M_{\odot}$ yr$^{-1}$) & \\
     \hline 
    A & $9.21 \pm 0.08$ & $(3.8 \pm 0.02) \times 10^{-2}$  & -10.64 & $37.97^{+0.7}_{-0.6}$ & $40.1^{+0.2}_{- 0.1}$ & 0.12 & $4.5 \times 10^{-3}$ & $3.7 \times 10^{-5}$\\
    B & $10.38 \pm 0.4$ & $(2.7 \pm 0.8) \times 10^{-1}$ & -10.94 &  $39.18_{-0.3}^{+0.3}$ & $41.1 ^{+0.1}_{-0.1}$ & 0.20 & $5.4 \times 10^{-2}$ & $1.8 \times 10^{-4}$ \\
    C & $8.83 \pm 0.3$ & $1.8 \pm 0.09$  & -8.57 & $39.75_{-0.5}^{+0.4}$ & $40.3 ^{+0.2}_{-0.1}$ & 0.10 & 0.18 & $2.8 \times 10^{-2}$\\
    \hline 
    \end{tabular}
\end{table*}

\subsection{Emission from X-ray binaries}
\label{sec: binaries}
In addition to AGN activity, another potential source of X-ray emission in galaxies is the integrated emission from X-ray binaries (XRB). To confirm that the X-rays from these sources are dominated by AGN, we calculate the luminosity (\lxb) that we would expect from galaxy-wide XRBs. Chandra studies of nearby star-forming late-type galaxies and passive early-type galaxies have shown that the X-ray point-source emission from relatively young high-mass XRBs and older low-mass XRBs correlate well with galaxy SFR and \mstar, respectively \citep[e.g.,][]{Grimm2003, Gilfanov2004, Lehmer2010, Lehmer2019}. Since the stellar masses listed in D21 are systematically underestimated for local galaxies, we use \mstar \, values derived from the SED in \S \ref{sec: sed}. These values are listed in Table \ref{tab: X-ray binary}.

\citet{Lehmer2019} present X-ray luminosity functions for binaries based on sub-galactic modelling. Following their Table 6, we estimate \lxb \, for our sources. We utilize the SFRs and stellar masses provided by the SED fits. For the former, we choose the SFR at 100 Myr since this better encapsulates the actively star-forming population over the timescales relevant for high-mass XRB formation, reducing stochastic effects from short-term variations in star formation \citep[e.g.,][]{Garofali2018, Lehmer2024}. Source A has a log mass of 9.21 $M_{\odot}$ and SFR of $3.8 \times 10^{-2}$, giving us log sSFR = -10.64\footnote{Due to the underestimation in FIR, a higher SFR would lead to a larger log sSFR. For example, an increase of 0.2 dex in log sSFR would lead to 0.03 dex increase in \lxb.}. Source B has log mass 10.38 $M_{\odot}$ and SFR of $2.7 \times 10^{-1}$ with log sSFR of -10.94. Source C has the lowest log mass of 8.83, SFR 1.8, with log sSFR = -8.5. These values are listed in Table \ref{tab: X-ray binary}; interpolating the data from Table 6 of \citet{Lehmer2019}, we use these to obtain estimates of log \lxb \, as $37.97^{+0.7}_{-0.6}, 39.18_{-0.3}^{+0.3}, 39.75^{+0.4}_{-0.5}$ for Sources A, B, and C, respectively. The emission from X-ray binaries alone cannot account for the observed emission in our sources, particularly in sources A and B, strongly suggesting the presence of AGN activity. Source C's \lxb \, differs from the observed \lx \, by 0.6 dex suggesting that the integrated emission from the galaxy could be powered by a population of X-ray binaries.

To remain consistent with the methodology outlined in \citet{Lehmer2019}, we estimate stellar masses and SFRs using the empirical relations related to optical luminosity and colour provided in \citet{Zibetti2009} and \citet{Hao2011} (see \S 3 in \citealt{Lehmer2019} for details). Our analysis reveals that the stellar mass estimates are systematically lower, while the SFRs are approximately an order of magnitude higher compared to values derived from SED fitting. However, upon interpolating from Table 6 of \citet{Lehmer2019}, we find that the \lxb \, values are slightly reduced relative to our previous estimates. The uncertainties remain consistent, ensuring agreement within the statistical errors.

A further constraint on the expected contribution from X-ray binaries comes from the fact that the observed X-ray emission for all three systems is unresolved and centred on the nucleus of the galaxy, which is expected to encompass only a fraction of the total X-ray binary emission. We estimate the SFR within the PU (vs the integrated galaxy emission). We assume the spatial distribution of SFR is traced approximately by the $u$-band flux, and used the Aperture Photometry Tool \citep[APT; ][]{Laher2012} to determine the fraction of the $u$-band flux contained within the 95\% confidence interval for the X-ray position. For each of the three sources, this nuclear $u$-band light fraction is $10-20$\%, so that the expected integrated luminosity from X-ray binaries is 5--10$\times$ smaller than the estimates above. 

\begin{figure}
    \centering
    \includegraphics[width=\columnwidth]{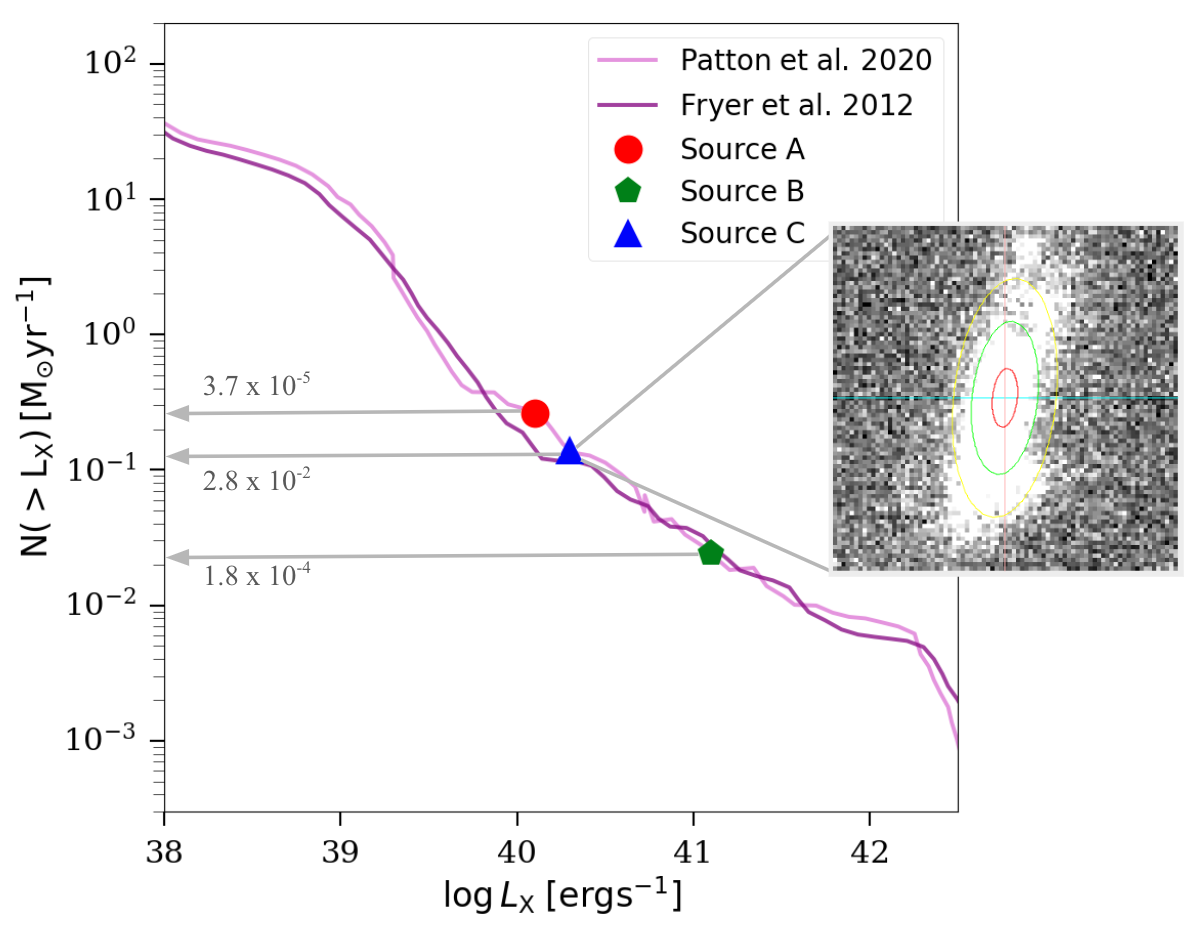}
    \caption{Relationships between the number of X-ray binaries in a galaxy per unit star formation rate, as a function of $L_X$. The values in grey represent the predicted number of XRBs with the observed luminosity. In the right panel, we show the aperture photometry done for source C to calculate the fraction of light enclosed within the PU.}
    \label{fig: xlf}
\end{figure}

We can further estimate the likelihood that star formation processes would produce a {\em single} X-ray binary with the observed luminosity, consistent with the fact that the sources are unresolved. (We note that Source B is definitively a single source, as demonstrated by its X-ray variability; \S \ref{sec: var}.) We utilize the synthetic population models studied in \citet{Misra2023} which probes different aspects of XRB evolution using models from \citet{Patton2020} and \citet{Fryer2012}. For each of the three sources, we check the likelihood that the observed X-ray luminosity is produced by a single source, given the SFR within the source aperture. In Figure \ref{fig: xlf}, we plot the models from \citet{Patton2020} and \citet{Fryer2012}, each corresponding to different physical parameters such as remnant mass, natal kick, and common-envelope efficiency. The probability of single XRB associated with star formation within the source aperture having an $L_X$ greater than the observed value is $3.7\times10^{-5}$, $1.8 \times10^{-4}$, and $2.8\times10^{-2}$ for Sources A, B, and C, respectively. 

Finally, we note that if the observed X-ray luminosity is from a single source, it would be classified as an ultraluminous X-ray source \citep[ULX; e.g.,][]{Kaaret2017, King2023}. Given the high X-ray luminosity, if the sources are not BHs, the ULXs would have to be powered by super-Eddington accretors \citep{Singha2024}. A constraint on this possibility comes from the velocity dispersions of emission lines, since super-Eddington accretion is associated with a strong outflow which is generated at the accretion disk surface via radiation pressure \citep{Shakura1973}. \citet{Fabrika2015} showed that known super-Eddington accretors have strong correlations between the He II $\lambda 4686$ and H$\alpha$ line widths. All of our sources have a significantly lower $\sigma_{\rm{H \alpha}}$ than $\sigma_{\rm{He II \lambda 4686}}$. This is an important deviation from the 1:1 expected correlation as H$\alpha$ is known to trace ionized and gas-shocked winds. Thus, it is apparent that the sources are not powered by super-Eddington accretors.

The accompanying analysis presented, in particular the evidence for MIR emission from the AGN from the SED fit, the position of the source at the centre of the galaxy, and a compact nuclear region, points toward the sources being AGNs.

\subsection{X-ray variability}
\label{sec: var}

Chandra observations of source B reveal that it is variable in the X-ray. We extract the light curves and the fluxes using Chandra Interactive Analysis of Observations \citep[CIAO; ][]{CIAO}. We use a 1-D power law to fit the X-ray spectra for 3 observations: 4221 (PI: Jones), 13134 (PI: Murray), and 18475 (PI: Kraft). For these observations, the data is re-processed using \code{chandra\_repro} to discard bad pixels. With \code{specextract}, we extract the spectra of the source and generate the necessary Auxiliary response files and redistribution matrix files. With \textit{Sherpa}, we generate a 1D power-law model fit to source counts spectrum and obtain residuals.

\begin{figure}
    \centering
    \includegraphics[width=\columnwidth]{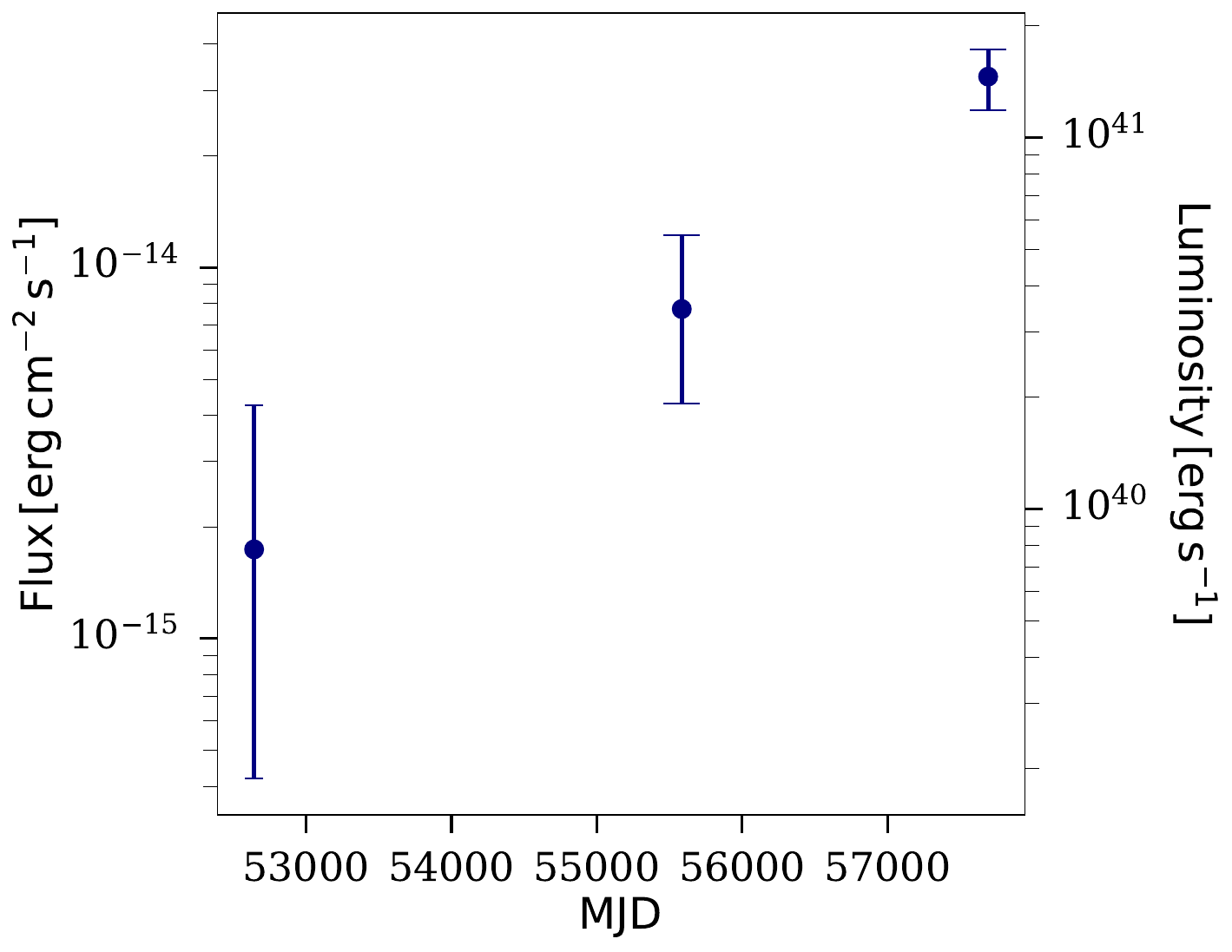}
    \caption{X-ray light curve in the 0.5--7 keV band for source B, created with fluxes calculated using CIAO tools. The three data points correspond to three observations for the source (from left to right: 4221, 13134, 18475).}
    \label{fig: xray_lc}
\end{figure}

To get a better sense of this variability, we calculate the flux in the three different available observations using \code{dmextract}. These reveal that we observe the source during its on-and-off phases. In the high flux phase, it has a flux of $3.3_{-0.6}^{+0.6} \times 10^{-14}$ erg cm$^{-2}$ s$^{-1}$. In the low flux phase, the flux is $1.7_{-2.5}^{+1.3} \times 10^{-15}$ erg cm$^{-2}$ s$^{-1}$. For Observation 1334, the flux is $7.7_{-3.4}^{+4.4} \times 10^{-15}$ erg cm$^{-2}$ s$^{-1}$. The luminosities associated with observations 18475, 4221, and 13134 are $1.4_{-0.2}^{+0.2} \times 10^{41}$, $7.7_{-1.8}^{+3.4} \times 10^{39}$, $3.4_{-1.5}^{+2.0} \times 10^{40}$ in erg s$^{-1}$. The light curve generated by these is shown in Figure \ref{fig: xray_lc} and the fluxes are listed in Table \ref{tab: Flux}. This variability by an order of magnitude suggests that the luminous X-ray emission from this galaxy is dominated by a single source, and is consistent with variability seen in AGN \citep[e.g.,]{Matthews1963, Elliot1974, Mushotzky1993, Zhang2011}.

Sources A and C do not have enough counts in their observations to explore a spectral fitting and thus their long-term variability cannot be explored.

\begin{table}
\centering
\caption{X-ray fluxes in the 0.5--7 keV band for each observation for Source B, obtained using CIAO tools, with \code{specextract}, \code{dmextract}, and \textit{Sherpa}.}
\label{tab: Flux}
    \begin{tabular}{cc}
    \hline 
    Observation ID & Flux \\
      & (erg cm$^{-2}$ s$^{-1}$) \\
      \hline
    4221 & $1.7_{-2.5}^{+1.3} \times 10^{-15}$ \\
    13134 & $7.7_{-3.4}^{+4.4} \times 10^{-15}$ \\
    18475 & $3.3_{-0.6}^{+0.6} \times 10^{-14}$ \\
    \hline 
    \end{tabular}
\end{table}

\section{Black Hole Mass Estimates}
\label{sec:mass_estimates}
In this section, we estimate the masses of the BHs (\mbh) using the \msigma \, and \mmstar \,  scaling relations and estimate the growth timescales via the observed Eddington ratios. Evidence points towards the scaling of central BH properties with their host galaxies properties like the stellar velocity dispersion, mass, IR luminosity, etc. The correlations provide an opportunity to study the linked evolution of galaxies and BHs and simultaneously provide a way to estimate the BH mass via a proxy \citep{Reines2015}. 

\subsection{\mmstar}

The scaling relation between the BH mass and \mstar \; has been studied in detail \citep[e.g.,][]{KormendyHo2013, Reines2015, Greene20}. We use the relations derived from \citet{Reines2015} and \citet{Greene20} to estimate \mbh. \citet{Reines2015} derive the total galaxy stellar mass using photometry \citep{Zibetti2009}. We use the masses calculated in \S \ref{sec: binaries} for \mstar, and estimate \mbh, from \citet{Reines2015}:

\begin{equation}
    \log{\left(\frac{M_{\text{BH}}}{M_{\odot}}\right)} = \alpha + \beta \log{\left(\frac{M_{\star}}{10^{11} M_{\odot}} \right)}
\end{equation}

where $\alpha = 7.45 \pm 0.08$ and $\beta = 1.05 \pm 0.11$. \citet{Greene20} presents another equation for the \mmstar \ scaling relation derived mainly using dynamical masses:

\begin{equation}
    \log{\left(\frac{M_{\text{BH}}}{M_{\odot}}\right)} = \alpha + \beta \log{\left(\frac{M_{\star}}{3 \times 10^{10} M_{\odot}} \right)} + \epsilon
\end{equation}

where $\alpha =  7.56 \pm 0.09$, $\beta =  1.39 \pm 0.13$, and $\epsilon = 0.79 \pm 0.05$. Focusing on the low-mass regime, it is apparent that without upper limits, the fit to late-type galaxies returns a very shallow relation because the measured \mbh \, values are biased towards SMBHs \citep[e.g.,][]{Pacucci2018, Baldassare2020a, Greene20}. The values are listed in Table \ref{tab: mass}. 

\subsection{\msigma}

\begin{figure}
    \centering
    \includegraphics[width=\columnwidth]{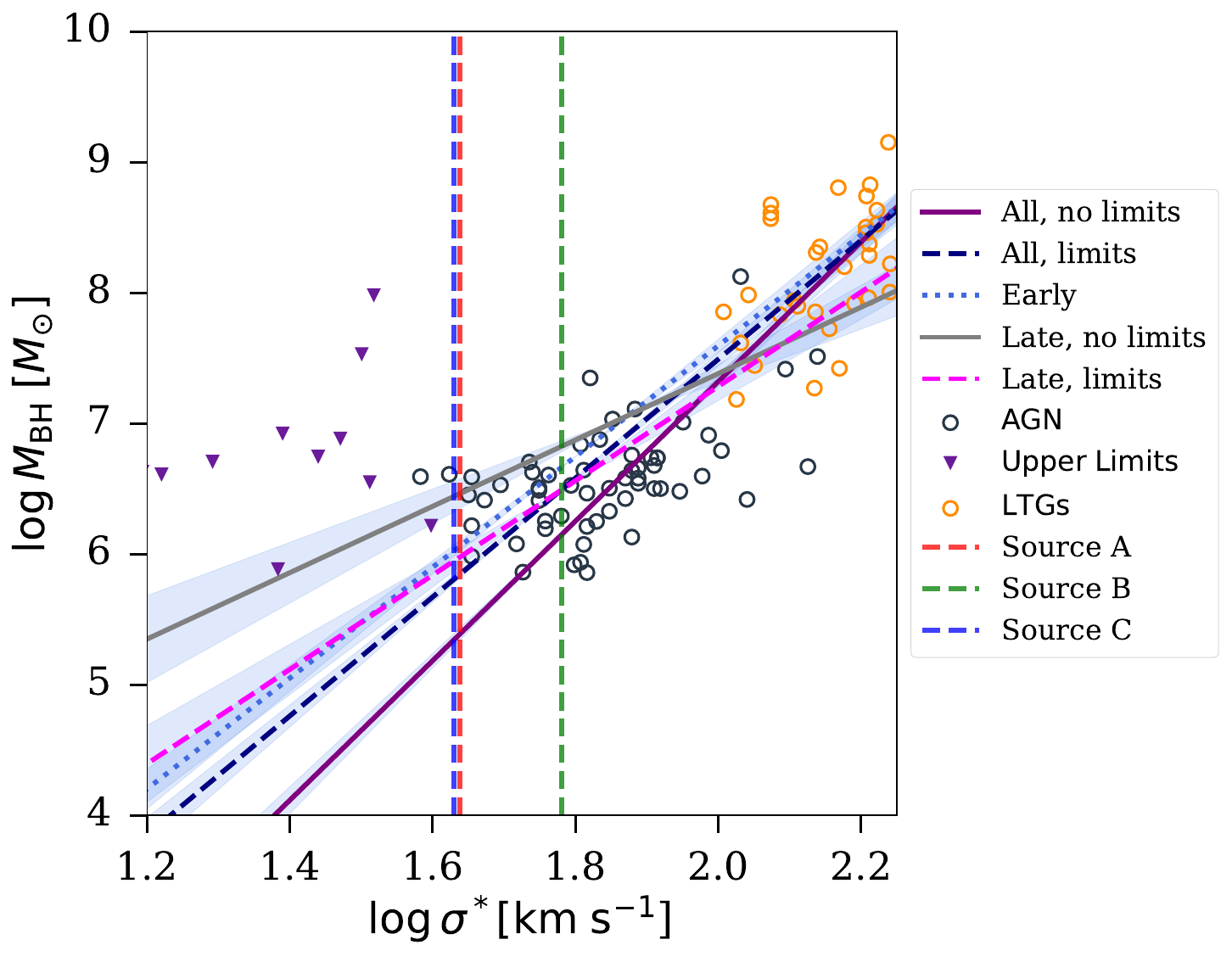}
    \caption{Sources on the \msigma \, relation following \citet{Greene20}. The gray open circles represent the AGN sample \citep{Xiao2011}, open yellow circles are late-type galaxies, and solid purple triangles are the upper limits on dwarf galaxies. The \msigma \, relationship shows a large scatter at the low mass tail which is in part due to the lack of uniform occupation fraction of BHs at lower masses. The estimated \mbh \, values of the sources are shown as shaded regions to convey related errors.}
    \label{fig: m-sigma}
\end{figure}

The \msigma \, relation is an empirical correlation between the BH mass and the stellar velocity dispersion of the host galaxy \citep[\vel; e.g.,][]{KormendyHo2013, Saglia2016, 2016ApJVanden, Krajnovi2018}. We follow the relation in \citet{Greene20} that is derived using BH mass samples from \citet{KormendyHo2013}, \citet{Greene2016}, \citet{Krajnovi2018}, \citet{Nguyen2018} and \citet{Thater2019}:

\begin{equation}
    \left(\frac{M_{\text{BH}}}{M_{\odot}} \right) = \alpha + \beta \log{\left(\frac{\sigma^*}{160 \; \text{km/s}}\right)} + \epsilon ,
\end{equation}

where $\alpha = 7.88 \ \pm \ 0.05$, $\beta = 4.34 \, \pm \ 0.24$, and $\epsilon = 0.53 \, \pm \, 0.04$. The estimates we obtain for the masses are shown in Figure \ref{fig: m-sigma} and tabulated in Table \ref{tab: mass} for ``All, no limit" relationship of \citet{Greene20}. 

The scatter in these relationships has been widely studied \citep[e.g.,][]{KormendyHo2013, Saglia2016, Pacucci2018} and is primarily due to the uncertainties in measurements of galaxy properties and due to the presence of only upper limits on some quantities. These values can be found in Table \ref{tab: mass}.

\begin{table*}
\caption{A tabulation of the BH mass estimates and Eddington ratios calculated using various mass scaling relationships}
\label{tab: mass}
    \begin{tabular}{ccccccc}
    \hline 
    Identifier & \vel & \msigma & \mmstar (1) & \mmstar (2) & log $L_{\rm Edd}$ & $\lambda_{\rm Edd}$ \\
     &  (km s$^{-1}$) & log\mbh \, ($M_{\odot}$) & log\mbh \, ($M_{\odot}$) & log\mbh \, ($M_{\odot}$) & (erg s$^{-1}$) & \\
    \hline 
    A & $43.5 \pm 7.6$ & $5.42 \pm 0.33$ & $5.58$ & $5.80$ & $43.52 $ &  $0.02$ \\
    B & $60.3 \pm 12.5$ & $6.04 \pm 0.40$ & $6.85$ & $7.49$ & $44.13$ & $0.13$ \\
    C & $42.6 \pm 10.2$ & $5.38 \pm 0.57$ & $5.17$ & $5.27$ & $43.48$ & $ 0.39$ \\
    \hline
    \end{tabular}
\end{table*}

\subsection{Discussion of scaling relations}

We choose to rely on the \msigma \, derived BH masses as our main findings due to the large scatter in the \mmstar \, relation, and because of the significant deviation between $\sigma$ and \mstar-derived masses for source B. 

The SDSS image of source B strongly suggests a disk morphology, with a diameter of $\sim 30"$. The stellar velocity dispersion is well below what is expected for a spheroid-dominated system of similar \mstar \, \citep{Leigh2012}, indicating that its stellar morphology is dominated by a disk. However, we note that the SDSS Legacy fiber aperture is 3" \citep{Adelman2007} and that the measured velocity dispersion may still be dominated by a bulge component.

We should note that the applicability of the \msigma \, relation is debated for low-mass galaxies \citep[e.g.,][]{Pacucci2018, Greene20, Reines2022} because \vel \, is traditionally defined as the velocity dispersion of stars inside bulges. The slope of the relation changes considerably based on the galaxy population, likely due to the bias in \mbh \, measurements toward the most massive BHs. There is also a marked departure in the estimates for $M_{\text{BH}} \leq 10^5 M_{\odot}$ \citep{Pacucci2018}. The observed \msigma \, relation favours heavy seed models that make exclusively \mbh \ $ > 10^5 M_{\odot}$. The low-mass tail of the \msigma\ relation is now being explored since it is believed to provide a stronger, and more fundamental, correlation than \mmstar \, \citep[e.g.,][]{Sesana2016, Volonteri2010, Pacucci2018}.

SDSS recommends users not rely upon data with \vel $< 70$ km s$^{-1}$ due to the typical signal-to-noise ratio and the instrumental resolution of the spectrograph\footnote{https://classic.sdss.org/dr5/algorithms/veldisp.php} \citep[for more information, see][]{Adelman2007, Almeida2023}. The instrumental dispersion of the SDSS spectrograph is 69 km s$^{-1}$ per pixel, the resolution is $\sim90$  km s$^{-1}$, and velocities are convolved to a maximum value of 420 km s$^{-1}$. Since our sources are low-mass and hence have small \vel, their velocity dispersions fall below the recommended limit. To verify the validity of our estimate, we carried out our analysis of the SDSS spectra from which the velocity dispersions were derived. We confirmed that the \vel \, measurements are reliable in terms of the strength of the relevant absorption lines, the spectral resolution of the data, and the model fit to the flux to determine the velocity dispersion. We first fit Gaussians to the calcium absorption and neon emission lines to check the \vel \, derived from the fit width against that reported by SDSS. 

For a more rigorous analysis, we use the \code{ppxf} Python package \citep{Cappellari2022} to carry out a detailed analysis using both photometric and spectroscopic data. Using stellar and gas templates, we fit a two-component model (V: velocity and \vel: velocity dispersion) to the data. We use the MILES Library of Stellar Spectra derived from \citet{Sanchez2006} and \citet{Falcon2011}. The fitting takes into account a parameter called \code{sigma\_diff} which is the difference between the instrumental dispersion of the galaxy spectrum and the instrumental dispersion of the template spectra\footnote{https://pypi.org/project/ppxf/}. This proves helpful since the templates are convolved using an analytic Fourier Transform by \code{ppxf} to match the provided instrumental dispersion of the galaxy spectrum, in this case, the instrumental resolution of the spectrograph. We find that the estimates agree with the SDSS values and fall within the reported SDSS errors. Thus, we adopt the \vel-derived BH mass in calculating the Eddington ratio for all three sources. The masses of the three sources are $\sim 10^5-10^6 M_{\odot}$ as listed in Table \ref{tab: mass}.

Mass scaling relations vary greatly with galaxy types and merger histories. \citet{Graham2023} presents scaling relations for E, S0, and S galaxies, emphasizing the impact of morphology and merger history. We studied these various scaling relations, the dataset from \citet{Sahu2019} used to derive them, and the velocity dispersions for our sources. We found that the \vel \, values of our sources place them on the low-mass tail of the relations. The estimated masses from these relations are consistently smaller than estimates from \citet{Greene20}. Median BH masses for the sources are $2.2 \times 10^4$ $M_\odot$ for A, $1.6 \times 10^5$  $M_\odot$ for B, and $2.0 \times 10^4$ $M_\odot$ for C.

With the estimated masses, we calculate the Eddington limits of the sources. To compute the corresponding Eddington ratio, we first find the bolometric luminosity. We use the 15 $\mu$m MIR bolometric luminosity relationship in \citet{Shen2020} to convert the observed luminosity at 15 $\mu$m to the corresponding bolometric luminosity. The observed MIR luminosity at 15 $\mu$m is extracted from the pure AGN component of the SED template (Figure \ref{fig: SEDs}). The bolometric luminosities are listed in Table \ref{tab: mass} and from these we calculate  Eddington ratios of $\sim 10^{-2} - 10^{-1}$. As these sources are radiating at a relatively high fraction of their Eddington limit, we can infer that it is currently in a significant phase of growth. The Eddington ratios are also consistent with thin-disk accretion and thus lie in the radiatively efficient regime of AGNs \citep{Shakura1973}.

\section{Conclusion}
\label{sec:conclusion}
In this paper, we analyse the nature of three low-mass galaxies with strong nuclear X-ray emission in the Bo\"otes field and investigate whether they host accreting nuclear BHs. These targets are selected by cross-matching two catalogues: \citet{Masini2020} (X-ray detections) and \citet{Duncan2021} (galaxies, with estimates of total stellar mass of total stellar mass). After matching with a 1" radius, we  made initial cuts to select nearby ($z < 0.05$), low-mass galaxies ($M_{\star} < 10^{9.5} M_{\odot}$). After making additional cuts on data requiring multi-wavelength coverage and appropriate velocity dispersions, we were left with three sources that we study in this paper. With SDSS data we analyse their emission lines by placing them on a narrow-line BPT diagnostic diagram. We calculate F-band X-ray luminosities and rely on Chandra detections to identify AGN because they are one of the clearest indicators of AGN activity in a galaxy \citep[e.g.,][]{Miller2015, Baldassare2015, Pardo2016}. Furthermore, we retrieve photometry in the UV, optical, and IR to compute the SEDs for our galaxies and fit them using CIGALE using various combinations of stellar and AGN templates. With the MIR and [O III] luminosities, we are able to assess the possibility that these galaxies host AGN. To confirm that the X-ray emissions we observe originate from AGNs and not galactic-wide X-ray binaries, we estimate the latter with SFRs. We find convincing evidence for our sources to host AGNs at their centres.

By analysing the deviation from intrinsic relationships between [O iii], MIR, and X-ray luminosities, we conclude that these sources are considerably obscured in the X-rays. Using the model presented in \citet{Carroll2023}, we estimate the column densities ($\log N_{\rm H}[{\rm {cm^{-2}}}] \sim 22.7, > 25.0$, and $24.4$) of the sources by considering the ratio of the observed to expected \lx. We use scaling relations (\msigma \ and \mmstar) to estimate the masses of the central BHs. The morphology of source B in Figure \ref{fig: images} and the $M_{\star} - \sigma^*$ relation trends for spheroids, we believe that it is a disc galaxy. Thus, we choose to use the \vel-derived relation for the BH masses for all three sources to be consistent. Their masses fall in the range of $10^5 - 10^6 M_{\odot}$. We estimate the Eddington ratios of the BHs using the \msigma-determined value for their masses, which are of the orders of $10^{-2} - 10^{-1}$. These ratios suggest that BHs are accreting at rates characteristic of optically-thick, radiatively efficient accretion flows.

We also find that one of our sources is highly variable (by approximately an order of magnitude) in the X-ray band. Using CIAO tools, we generate a light curve from three Chandra observations of the source. These observations offer us a view of the galaxy in its low-flux and high-flux phases, and confirms that the emission is dominated by a single source rather than a population of many X-ray binaries. Studying X-ray variability of AGNs in low-mass galaxies in the future will be important to understand the nature of their accretion disks, flows, and radiation mechanisms.

To place these low-mass galaxies in the context of the local Universe, it is helpful to consider the fraction of the low-mass galaxy population that we recover as AGNs. From \citet{Duncan2021}'s compilation, there are 66 galaxies in the Boötes field with $M_{\star} < 10^{9.5} M_{\odot}$ and spectroscopic redshift $z< 0.05$. Using the mass-to-light ratio method of estimating stellar masses as described in \S \ref{sec: binaries}, we note that stellar mass estimates of these sources are comparable to our three targets. Using these numbers, we can conclude that roughly 5\% of all local low-mass galaxies host AGNs. This is an important statistic in uncovering the presence of AGNs in local low-mass galaxies and finding more of these ``mini monsters" \citep{Hickox2022HEAD}. 

This is the first study of low-mass BHs and AGNs in low-mass galaxies in the Boötes field which has extensive multi-wavelength observations. In our work, we identified  targets using X-ray data archival Chandra data. More focused observations, spectra, and large fields of X-ray, optical, and infrared of local low-mass galaxies could help us to better identify low-mass AGNs. Additionally, a majority of the AGN activity in the universe is obscured \citep[e.g.,][]{Ueda2014, HickoxAlexander2018, Ananna2019}, meaning that studying obscured AGNs in low-mass galaxies can provide a more complete picture of the population fraction and properties of low-mass BHs. 


\section*{Acknowledgements}
    We thank the reviewer for their insightful comments that greatly improved this manuscript. R.A.P. acknowledges support from the James O. Freedman Presidential Scholarship at Dartmouth College. R.C.H. acknowledges support from NASA through grant number 80NSSC22K0862. G.C.P. acknowledges support from the Dartmouth Fellowship. We thank Christopher Carroll, Quinn Casey, Emmanuel Durodola, Jenny Greene, Stephanie Podjed, and Kelly Whalen for insightful discussions.
    
    Funding for SDSS has been provided by the Alfred P. Sloan Foundation, the Participating Institutions, the National Science Foundation, the U.S. Department of Energy, the National Aeronautics and Space Administration, the Japanese Monbukagakusho, the Max Planck Society, and the Higher Education Funding Council for England. This research has made use of the NASA/IPAC Extragalactic Database (NED), which is funded by the National Aeronautics and Space Administration and operated by the California Institute of Technology. Much of this work was carried out at Dartmouth College, Hanover, which stands on the ancestral lands of the Abenaki and Algonquin peoples.

\section*{Data Availability}
Publicly-available data was used in this project, obtained from NED, SDSS, and Chandra's WebChaser. All data products used in this paper will also be made available upon reasonable request to the authors.


\bibliographystyle{mnras}
\bibliography{bib}

\bsp	
\label{lastpage}
\end{document}